\newif\iffull\fulltrue
\newif\ifacm\acmfalse

\ifacm
    \documentclass[acmsmall,screen]{acmart}
\else
    \documentclass{extarticle}
\fi

\ifacm
    \acmPrice{}
    \acmDOI{10.1145/3498690}
    \acmYear{2022}
    \copyrightyear{2022}
    \acmSubmissionID{popl22main-p160-p}
    \acmJournal{PACMPL}
    \acmVolume{6}
    \acmNumber{POPL}
    \acmArticle{29}
    \acmMonth{1}
    \startPage{1}

    \setcopyright{rightsretained}

\else 

\fi

\ifacm 
\else 
    \usepackage{amsmath}
    \usepackage{amssymb}
    \usepackage{amsthm}
    \usepackage[numbers]{natbib}
    \usepackage[margin=3cm]{geometry}
\fi

\usepackage[margin=false,inline=true]{fixme}
\FXRegisterAuthor{aaa}{anaaa}{\color{cyan}AAA}
\FXRegisterAuthor{mg}{anmg}{\color{red}MG}
\FXRegisterAuthor{cz}{ancz}{\color{orange}CZ}
\newcommand{\aaa}[1]{}
\newcommand{\mg}[1]{}
\newcommand{\cz}[1]{}

\usepackage{stmaryrd}

\usepackage{stackengine}
\usepackage{mathrsfs}
\usepackage{braket}

\usepackage{tikz}
\usetikzlibrary{shapes.geometric, positioning, arrows, matrix}

\usepackage{tikz-cd}

\usepackage{mathpartir}

\usepackage{hyperref}
\usepackage{cleveref}

\usepackage{enumitem}

\usepackage{listings}

\usepackage{booktabs}

\usepackage{subcaption} 

\iffull
    \usepackage[conf={restate}]{proof-at-the-end}
\else
    \usepackage[conf={restate, no link to proof}]{proof-at-the-end}
\fi

\newtheorem{definition}{Definition}
\newtheorem*{definition*}{Definition}
\newtheorem{example}{Example}

\newcommand{\bool}{\mathbb{B}}
\newcommand{\true}{\mathrm{true}}

\newcommand{\powerSet}[1]{{\mathcal{P}(#1)}}
\newcommand{\limplies}{\to}

\newcommand{\codomain}[1]{\mathrm{cod}(#1)}
\DeclareMathOperator{\post}{\mathrm{post}}

\newcommand{\thenCompo}{\mathbin{\fatsemi}}

\newcommand{\union}{\cup}
\newcommand{\bigunion}{\bigcup}
\newcommand{\compl}[1]{\overline{#1}}
\newcommand{\reverse}[1]{{#1}^{\vee}}

\newcommand{\validUnderVal}[2]{\models_{#1} {#2}}
\newcommand{\validUnderModels}[2]{{#1} \models {#2}}

\newcommand{\interpWith}[2]{{\sem{#2}_{#1}}}

\newcommand{\nat}{\mathbb{N}}

\newcommand{\starOf}[1]{{#1}^{\star}}
\newcommand{\fail}{\mathinner{\mathtt{Fail}}}

\newcommand{\sem}[1]{\llbracket #1 \rrbracket}

\newcommand{\definedAs}{\triangleq}

\newcommand{\okState}{\mathinner{ok}}
\newcommand{\errState}{\mathinner{er}}

\newcommand{\falsehood}{\bot}

\newcommand{\command}[1]{\mathtt{#1}}
\newcommand{\comSkip}{\command{skip}}

\newcommand{\comError}{\command{error ()}}
\newcommand{\comAssume}[1]{\command{assume}~#1}
\newcommand{\comITE}[3]{\command{if}~#1~\command{then}~#2~\command{else}~#3}
\newcommand{\comWhile}[2]{\command{while}~#1~\command{do}~#2}
\newcommand{\comAssign}[2]{#1:= #2}

\newcommand{\comIter}[1]{{#1}^{\star}}

\newcommand{\incorTriple}[3]{[#1]~#2~[#3]}
\newcommand{\hoareTriple}[3]{\{#1\}~#2~\{#3\}}

\newcommand{\tand}{\text{ and }}

\newcommand{\twhere}{\text{ where }}
\newcommand{\tst}{\text{ s.t.\ }}

\newcommand{\itemTitle}[1]{\textbf{#1}}
\newcommand{\keyword}[1]{\emph{#1}}

\newcommand{\algebra}[1]{\mathcal{\MakeUppercase{#1}}}
\newcommand{\modelsOf}[1]{\ensuremath{\mathsf{\mathbf{#1}}}}
\newcommand{\synt}[1]{\ensuremath{{\mathsf{#1}}}}

\newcommand{\languageModel}{{\algebra{G}}}
\newcommand{\guardedTerm}{{GT}}
\newcommand{\topLangModel}{{\algebra{G}_{\top}}}
\newcommand{\topGuardedTerm}{{\guardedTerm_{\top}}}
\newcommand{\unitTopLangModel}{{1_{\topLangModel}}}
\newcommand{\TopKATStar}{\ensuremath{\text{TopKAT}^{\star}}}

\newcommand{\allKATs}{\texorpdfstring{\modelsOf{KATs}}{KATs}}
\newcommand{\allRELs}{\texorpdfstring{\modelsOf{RELs}}{RELs}}
\newcommand{\allTopRELs}{\texorpdfstring{\modelsOf{TopRELs}}{TopRELs}}
\newcommand{\allFailTopRELs}{\texorpdfstring{\modelsOf{FTopRELs}}{FTopRELs}}
\newcommand{\allGRELs}{\texorpdfstring{\modelsOf{TopGRELs}}{TopGRELs}}
\newcommand{\allTopKATs}{\texorpdfstring{\modelsOf{TopKATs}}{TopKATs}}

\newcommand{\allFailTopKATs}{\texorpdfstring{\modelsOf{FailTopKATs}}{FailTopKATs}}
\newcommand{\KATTerm}[2]{\mathrm{KATTerm}_{#1, #2}}
\newcommand{\TKATTerm}[2]{\mathrm{TopKATTerm}_{#1, #2}}

\begin{document}

\title{On Incorrectness Logic and Kleene Algebra with Top and Tests}         

\ifacm
\else
    \author{
        Cheng Zhang \\
        \and
        Arthur Azevedo de Amorim \\
        \and
        Marco Gaboardi \\
    }
    \date{}
\fi

\ifacm 
\else 
    \maketitle
\fi

\begin{abstract}
  Kleene algebra with tests (KAT) is a foundational equational framework for
  reasoning about programs, which has found applications in program
  transformations, networking and compiler optimizations, among many other
  areas. In his seminal work, \citeauthor{kozen_hoare_2000} proved that KAT
  subsumes propositional Hoare logic, showing that one can reason about the
  (partial) correctness of while programs by means of the equational theory of
  KAT\@.
  In this work, we investigate the support that KAT provides for reasoning about
  \emph{incorrectness}, instead, as embodied by
  \citeauthor{ohearn_incorrectness_2020}'s recently proposed incorrectness
  logic.  We show that KAT cannot directly express incorrectness logic.  The
  main reason for this limitation can be traced to the fact that KAT cannot
  express explicitly the notion of codomain, which is essential to express
  incorrectness triples. To address this issue, we study Kleene Algebra with
  Top and Tests (TopKAT), an extension of KAT with a top element. We show that
  TopKAT is powerful enough to express a codomain operation, to express
  incorrectness triples, and to prove all the rules of incorrectness logic
  sound. This shows that one can reason about the incorrectness of while-like
  programs by means of the equational theory of TopKAT\@.
  \mg{I think we should be more precise in the abstract. we don't talk
    about relational models but most of the results are about
    relational models. Let's revise this after the intro.}
\end{abstract}

\section{Introduction}


Since the seminal work of \citet{Floyd67} and \citet{Hoare69}, logic has become
an essential tool for program verification.  A program logic provides a system
of \emph{deduction rules} to prove \emph{Hoare triples} \(\hoareTriple{b}{p}{c}\),
where \(p\) is a program, and \(b\) and \(c\) are assertions describing the pre-
and post-conditions.  Such Hoare triples are (partial) correctness
specifications, which state that, if we run \(p\) on an initial state satisfying
\(b\), all the final states (if reached) will satisfy \(c\). In other words, \(c\)
\emph{over-approximates} the set of final states of \(p\) starting from \(b\).

Despite their popularity, such logics are not the only tool for verifying
programs. Another long-standing approach, which can be traced back to works by
\citet{Taylor79} and \citet{odonnell_1985} in the 80s, is
\emph{equational reasoning}. Programs are modeled as elements of some algebraic
structure, whose equational theory can be used to prove specifications. One such
algebraic structure is \emph{Kleene algebra with tests}~\cite{kozen_kleene_1997}
(KAT), which extends Kleene algebra with predicates for modeling
conditionals. Kleene algebras with tests have several pleasing properties, such
as equality of terms being decidable in PSPACE, and have been applied in several
domains, including program transformations~\cite{angus_kleene_2001}, 
networking~\cite{anderson_netkat_2014,smolka_cantor_2017}, 
compiler optimization~\cite{kozen_certification_2000}, and more.

The two approaches turned out to be not so different. \citet{kozen_hoare_2000}
showed that KAT can express the validity of a Hoare triple as an equation, in
such a way that the deduction rules of a large fragment of Hoare Logic can be
obtained by equational reasoning.  \citeauthor{kozen_hoare_2000}'s work
established a clear bridge between deductive and equational reasoning for
program verification, thus shedding light on the expressive power of KAT\@.

In this work, we are interested in extending this correspondence to other
deductive formalisms for reasoning about programs.  One such formalism is
\emph{incorrectness logic} (IL), a recent proposal
by~\citet{ohearn_incorrectness_2020} that relates to earlier works
by~\citet{barthe_reverse_2011}. 
Instead of correctness, as in the Floyd-Hoare tradition, the system
revolves around \emph{incorrectness specifications} of the form
\(\incorTriple{b}{p}{c}\), which state that \(p\) can produce \emph{any} final
state satisfying \(c\) from \emph{some} input state satisfying \(b\)---or,
equivalently, that \(c\) \emph{under-approximates} the set of final states of
\(p\) starting from \(b\).  Though less conventional than traditional Hoare
logic, incorrectness logic has already found its way to several applications,
such as variants of separation logic~\cite{RaadBDDOV20} 
and relational verification for noninterference~\cite{Murray_2020}.

\begin{figure}
    \centering
    \begin{tikzcd}[row sep=5em, column sep=10em, math mode = false]
        \allFailTopRELs \arrow[r, two heads, 
            "\Cref{the: FailTopKAT can express incorrectness logic}"]
        & \text{IL} \\
        \allTopRELs \arrow[r, two heads, 
            "\Cref{the: topkat can express incorrectness logic normal termination}"]
        & \text{codomain inclusion} 
            \arrow[d, "defines"description]
            \arrow[dd, bend left=90, "defines"description]\\
        \allGRELs \arrow[d, leftrightarrow, "\Cref{the: GREL expresses same predicate as REL KAT} (equiexpressive)"' ]
            \arrow[r, "/" marking, two heads, "\Cref{the: GREL cannot formulate incorrectness logic}"]
        &  \text{IL (without errors)} \\
        \allRELs \arrow[r, two heads, "\cite{kozen_hoare_2000}"']
            \arrow[ur, "/" marking, two heads, "\Cref{the: KAT not able to express incorrectness logic}"']
        & \text{HL } 
    \end{tikzcd}
    \caption{Expressiveness of different systems.}\label{fig:expressiveness-of-different-kat}
\end{figure}
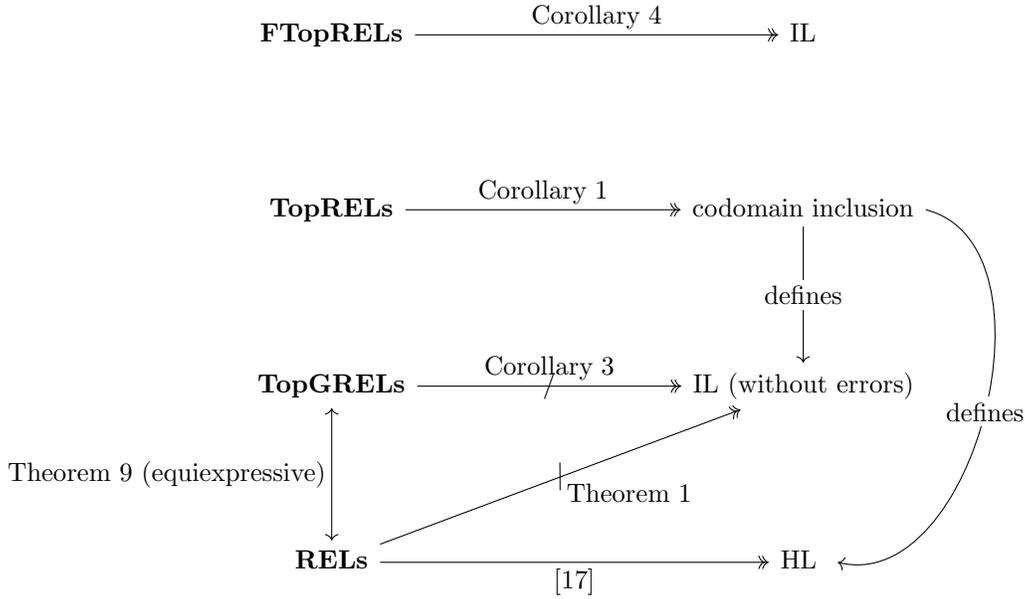

It is natural to wonder whether \citeauthor{kozen_hoare_2000}'s idea could be
adapted to encode incorrectness logic in Kleene algebra with tests.
Unfortunately, this is not the case. As we will show in this paper, there are
incorrectness triples that cannot be expressed by any KAT equation.  This might
appear surprising, given the symmetry between over- and under-approximation in
the formulation of Hoare logic and incorrectness logic.  However, the symmetry
involves the \emph{image} of a set by a relation, an operation that is not part
of the syntax of KAT\@.  Several prior works have considered enlarging KAT with
similar
operations~\cite{desharnais_modal_2004,fahrenberg_domain_2021,desharnais_kleene_2006},
but we show here that a smaller extension also serves our purposes: namely,
adding a top element \(\top\) to KAT\@.   We call such a structure a \emph{Kleene
  algebra with tests and top}, or a TopKAT\@.  We show that such structures can
encode inequalities between images, which we use to express incorrectness
triples.  The encoding allows us to prove the rules of incorrectness logic
equationally, thus extending \citeauthor{kozen_hoare_2000}'s correspondence to
incorrectness reasoning. In fact, the use of TopKAT to encode incorrectness logic was also suggested by~\citet{ohearn_incorrectness_2020}.

For clarity of exposition, our main focus is on the fragment of incorrectness
logic that handles normal program termination.  However,
\citet{ohearn_incorrectness_2020} also considered triples of the form
\([p]\, c\, [er: q]\), whose interpretation is similar to the one we described
above, except that they assume that execution can terminate with a fatal error
(e.g.\ a failed assertion).  Following \citet{esparza_equational_2017}, we show
that our encoding carries over to such triples by considering FailTopKAT, an
extension of TopKAT that includes an element \(\fail\) for representing failure.
We prove that the abnormal termination rules of incorrectness logic follow from
the equations of FailTopKAT\@.

We summarize our encodings in \Cref{fig:expressiveness-of-different-kat}.  We
use the notation \(A \twoheadrightarrow B\) to mean that the logic \(B\) can be
expressed in the equational theory of \(A\).  More formally, we model the
ground-truth notion of validity in each logic as a statement about sets
(assertions about program states) and relations (the input-output graph of a
program).  The encodings show that such statements are equivalent to equations
involving operations in \emph{relational} algebraic structures, where the
carrier of the structure is some set of relations between program states (for
example, \(\allFailTopRELs\) is the class of relational FailTopKATs in
\Cref{def:relational-failtopkat}).  Moreover, we prove that the equational
theory alone (i.e., not specialized to relations) suffices to derive the rules
of each logic.  We also use the arrow \(\to\) to denote the fact that 
systems capable of expressing codomain can express both Hoare logic and incorrectness logic.

\aaa{The explanation about the \(\to\) notation in
  \Cref{fig:expressiveness-of-different-kat} seems a bit ad hoc.}

\begin{figure}
    \centering
    \begin{tikzcd}[math mode = false, row sep=4em, column sep=9em]
        & \allTopRELs \arrow[d, "/" marking, "incomplete"'near start, 
        "\Cref{the: topkatstar incomplete over relational model}"near start]
        & \\
        \allGRELs \arrow[r, "complete", "\Cref{the: topkat complete}"']
        & \allTopKATs & 
        \text{language \allTopKATs} \arrow[l, "complete"', "\Cref{the: topkat complete}"] \\
        \allRELs \arrow[r, "complete", "\cite{goos_kleene_1997}"'] & 
        \allKATs & 
        \text{language \allKATs} \arrow[l, "complete"', "\cite{goos_kleene_1997}"]
    \end{tikzcd}
    \caption{Completeness relationships between classes of Kleene Algebras with Tests.}\label{fig:completeness}
\end{figure}
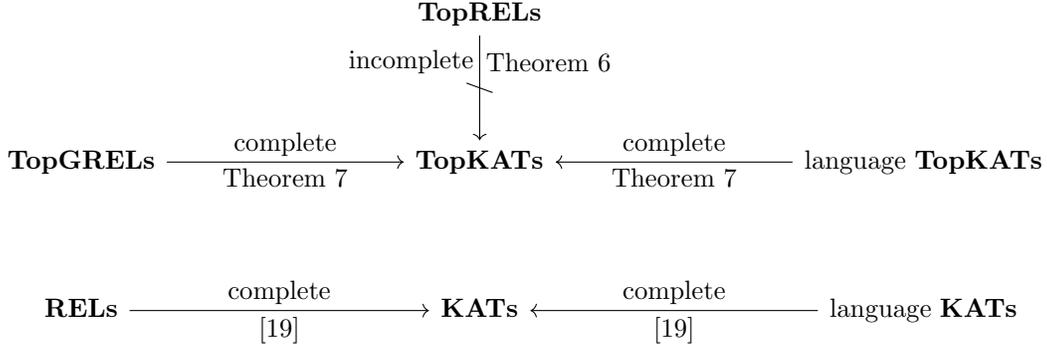

To evaluate the usefulness of these encodings, we investigate two basic
properties of TopKAT\@: \emph{completeness} and \emph{decidability}.  We say
that an equational theory is complete for a certain class of structures if it
can derive any equation that is valid in the class.  We are particularly
interested in completeness with respect to relational structures, since they are
the natural setting for formulating program logics. However, it is well-known that the
addition of a top element can be problematic for completeness of relational 
structures~\cite{pous_automata_2016}, 
and we show that this is the case for TopKAT as well: the theory is incomplete for
\(\allTopRELs\), the class of relational structures where the top element is the
complete relation.  However, we do get completeness by considering a larger
class \(\allGRELs\), where the top element might not be the complete relation.
We also show that TopKAT is complete for so-called \emph{language TopKATs}, a
class of structures inspired by prior work on KAT\@.
(Figure~\ref{fig:completeness} summarizes the relationships between these
different structures.)  Finally, we show that the equality of TopKAT terms can
be decided in PSPACE, by reducing a TopKAT term into a KAT term and 
applying the PSPACE algorithm for KAT equalities~\cite{cohen_complexity_1999}.

Summarizing, our contributions are:
\begin{itemize}
\item We show that (propositional) incorrectness logic cannot be encoded in
  relational KATs. Consequently, KAT cannot be used to reason equationally about
  incorrectness triples in general.
\item We consider TopKAT, an extension of KAT with an additional top element
  \(\top\), and show that (propositional) incorrectness logic for programs
  without error primitives can be encoded in relational TopKATs, by using \(\top\)
  to encode the codomain of a relation.  We prove that all the rules of this
  fragment of incorrectness logic can be derived solely by appealing to the
  equational theory of TopKAT\@.
\item We study the relations between the different systems we present
  in terms of expressivity and completeness.
\aaa{The phrasing sounds a bit weak. Can we relate this somehow to the original
  motivation in terms of incorrectness logic?}
\item We prove that deciding equality of TopKAT terms is PSPACE-complete.
\item We consider FailTopKAT, an extension of TopKAT by means of an element
  \(\fail\) and we show that this can be used to encode incorrectness logic with
  an error primitive.
\end{itemize}

Previous versions of this paper \cite{zhang_incorrectness_2022_arxiv,zhang_incorrectness_2022_POPL} 
contain an error in definition of language TopKAT
(Definition~13 and Lemma~1).
We have fixed this error in the current version, 
and this fix leads to a simpler proof of completeness, decidability, and complexity.
Now the proof of PSPACE-completeness and the decidability proof can be merged together,
and the section about \(\TopKATStar\) is no longer necessary;
hence we made changes to the section and theorem numbering,
but all the theorems in the old version are still provable.
We also want to acknowledge Damien Pous and Jana Wagemaker for pointing out the mistake.

\section{Background}

\subsection{Klenee Algebra with Tests}
Kleene algebra with tests was introduced by \citet{kozen_kleene_1997} as an
extension of Kleene algebra targeting program verification.  The equational
theory of Kleene algebras generalizes the one of regular expressions, and can be
used to reason about conditionals, loops, and simple (parametric) program
manipulations.



\begin{definition}[KAT]\label{def: KAT}
  A \emph{Kleene algebra} is an idempotent semi-ring \(\algebra{K}\) endowed with
  a Kleene star operation \(\starOf{(-)}\), satisfying the following properties:
  for all \(p, q, r \in \algebra{K}\):
\begin{align*}
    p + 0 = 0 + p & = p & \text{ identity} \\
    p + q &= q + p & \text{commutativity}\\
    (p + q) + r & = p + (q + r) & \text{associativity} \\
    p + p & = p & \text{idempotency} \\
    1 p = p 1 & = p & \text{identity} \\
    (p  q)  r & = p  (q  r) & \text{associativity}\\
    (p + q) r & = pr + qr & \text{right distributivity} \\
    r (p + q) & = rp + rq & \text{left distributivity} \\
    0 p = p 0 & = 0 & \text{annihilation} \\
    1 + \starOf{p}p = 1 + p\starOf{p} & = \starOf{p} 
    & \text{unfolding} \\
    q + pr \leq r & \implies \starOf{p}q \leq r & 
    \text{induction} \\
    q + rp \leq r & \implies q\starOf{p} \leq r & 
    \text{induction}, 
\end{align*} 
where the ordering \(\leq\) is defined as
\[
    p \leq q \iff p + q = q.
\]

A \emph{Kleene algebra with tests} (KAT, for short) is a pair
\((\algebra{K},\algebra{B})\), where \(\algebra{K}\) is a Kleene algebra of
\emph{actions} and \(\algebra{B} \subseteq \algebra{K}\) is a boolean
sub-algebra of \emph{tests}.  We call the class of all Kleene algebras with
tests \(\allKATs\).  We sometimes omit \(\algebra{B}\) if it can be inferred
from the context.  Tests are ranged over by \(a, b, c, d\), whereas
\emph{actions} are ranged over by \(p, q, r, s\).
\end{definition}

KATs can model program behavior by using actions to represent basic
components, tests to represent guards, multiplication to represent
sequential composition, addition to represent random choice, and star
to represent iteration. Concretely, \citet{kozen_kleene_1997} showed
that KATs can be used to model while programs using the following encoding:
\begin{align*}
    \comITE{b}{p}{q} & \definedAs b p + \compl{b} q \\
    \comWhile{b}{p} & \definedAs \starOf{(b p)} \compl{b}
\end{align*}

In the following sections, we will use KAT formulas to reason about the validity
of Hoare logic and incorrectness logic triples. Following
\citet{kozen_hoare_2000} we will see judgments in these logics as predicates
which can be expressed as KAT equalities. To do this, we will first need to
define KAT terms, their interpretation in a KAT, and what it means for a
predicate to be expressible using KATs.

We first need a notion of \emph{alphabet}, which is a pair \((K,B)\) of two
disjoint finite sets: an \keyword{action alphabet} \(K\) and a \keyword{test
  alphabet} \(B\).  We will refer to the elements of \(K\) as \keyword{primitive
  actions}, ranged over by \(\synt{p}, \synt{r}, \synt{q} \), similarly to
actions, and the elements of \(B\) as \keyword{primitive tests}, ranged over by
\(\synt{a}, \synt{b}, \synt{c}\), similarly to tests.

  We can now define the set of KAT terms.
  
\begin{definition}[KAT Terms]
  The set \(\KATTerm{K}{B}\) of \keyword{KAT terms} over the alphabet
  \((K, B)\) is generated by the following grammar:
\[\synt{t} \definedAs
  \synt{p} \in K \mid  \synt{b} \in B \mid 0 \mid 1 \mid
  \synt{t_1} + \synt{t_2} \mid
  \synt{t_1} \synt{t_2} \mid \starOf{\synt{t}} \mid \compl{\synt{t_b}},\] where
\(\synt{t_b}\) does not contain primitive actions.
\end{definition}




Terms can be interpreted using a valuation of the primitive actions
and tests in a KAT\@.

\begin{definition}[KAT Valuation and Interpretation]
  \label{def:kat-valuation}
  Let us consider  an alphabet \((K, B)\) and a KAT \((\algebra{K}, \algebra{B})\). A
  \keyword{valuation} is a function
  \(u: K \union B \to \algebra{K}\) such that \(u(\synt{b}) \in \algebra{B}\) for
  every \(\synt{b} \in B\).

  Given a valuation \(u\), we define the \keyword{interpretation}
  \(\interpWith{u}{-}: \KATTerm{K}{B} \to \algebra{K}\) as:
\begin{align*}
    \interpWith{u}{\synt{p}} &\definedAs 
    u(\synt{p}) & \text{if } \synt{p} \in K \union B \\
    \interpWith{u}{\synt{t_1} + \synt{t_2}} &\definedAs 
    \interpWith{u}{\synt{t_1}} + \interpWith{u}{\synt{t_2}} \\
    \interpWith{u}{\synt{t_1} \synt{t_2}} &\definedAs 
    \interpWith{u}{\synt{t_1}} \interpWith{u}{\synt{t_2}} \\
    \interpWith{u}{\starOf{\synt{t}}} &\definedAs 
    \starOf{\interpWith{u}{\synt{t}}} \\
    \interpWith{u}{\compl{\synt{t_b}}} &\definedAs 
    \compl{\interpWith{u}{\synt{t_b}}} &
    \text{if \(\synt{t_b}\) does not contain primitive actions}
\end{align*}
\end{definition}

Using the notion of interpretation, we can now define what it mean for an
equality between KAT terms to be \emph{valid}, which informally means that the
equality holds for every valuation.

\begin{definition}[Validity of KAT Equality]
  Given an alphabet \((K, B)\) and two KAT terms \(\synt{t_1}, \synt{t_2} \in \KATTerm{K}{B}\),
  a statement \(\synt{t_1} = \synt{t_2}\) is \keyword{valid under the valuation}
  \(u: K \union B \to \algebra{K}\) (denoted by
  \(\validUnderVal{u}{\synt{t_1} = \synt{t_2}}\)), if
\[\interpWith{u}{\synt{t_1}} = \interpWith{u}{\synt{t_2}}\]

A statement \(\synt{t_1} = \synt{t_2}\) is \keyword{valid} in all KATs, denoted as  \[\validUnderModels{\allKATs}{\synt{t_1} = \synt{t_2}},\] if 
\(\synt{t_1} = \synt{t_2}\) is valid under all KAT valuations.
\end{definition}

We can now state formally how we can use the equational theory of KATs
to reason about predicates.

\begin{definition}[Expressiveness of a KAT]
    Suppose that we have an alphabet \(K, B\), a KAT \(\algebra{K}\),
    an \(n\)-ary \keyword{predicate}  \(P: \algebra{K}^{n} \to \bool\)
    and \(n\) primitives \(\synt{p_1}, \dots, \synt{p_n} \in K \union B\).
    We say two terms \(\synt{t_1}, \synt{t_2} \in \KATTerm{K}{B}\)
    \keyword{express} the predicate \(P\)
    in \(\algebra{K}\) over \(\synt{p_1}, \dots, \synt{p_n}\),
    if for all valuations \(u: K \union B \to \algebra{K}\):
    \[\validUnderVal{u}{\synt{t_1} = \synt{t_2}} \iff 
    P(\interpWith{u}{\synt{p_1}}, \interpWith{u}{\synt{p_2}}, \dots \interpWith{u}{\synt{p_n}})\]
\end{definition}

Hoare logic and incorrectness logic treat programs as a relation between input
and output memories.  Accordingly, relation-based KATs will be fundamental to
formulate and manipulate these logics.

\begin{definition}[Relational KAT]
    A \keyword{relational KAT} \((\algebra{R}, \algebra{B})\) over a set \(X\) 
    is a KAT where
    \[\algebra{R} \subseteq \powerSet{X \times X}\]
    and tests \(\algebra{B} \subseteq \algebra{K}\), are subsets of identity relation on \(X\):
    \[\algebra{B} \subseteq \powerSet{\{(x, x) \mid x \in X\}}\]
    such that
    \begin{itemize}
        \item the addition operator \(+\) is the union of relations
        \item the multiplication operator is the sequential composition of relations:
        for \(p, q \in \algebra{R}\),
        \[pq = p \thenCompo q =
            \{(x, z) \mid \exists y \in X, (x, y) \in p, (y, z) \in q\}\]
        \item The additive identity 0 is the empty relation \(\emptyset\)
        \item The multiplicative identity 1 is the identity relation on \(X\):
        \[\{(x, x) \mid x \in X\}\]
        \item the star operator is the reflexive transitive closure:
        for \(p \in \algebra{R}\)
        \[\starOf{p} = \bigunion_{n \in \nat} p^{n}\]
        \item The complement of a test \(b \in \algebra{B}\) is:
        \[\compl{b} = 1 \setminus b\]
    \end{itemize}

    Some of the previous definitions can be extended to the relational setting:
    \begin{itemize}
        \item a \keyword{relational valuation} is a valuation in a relational KAT\@.
        \item a \keyword{relational interpretation} is an interpretation generated by a
        relational valuation
        \item a statement \(\synt{t_1} = \synt{t_2}\) is \keyword{relationally valid}
        (denoted \(\validUnderModels{\allRELs}{\synt{t_1} = \synt{t_2}}\)), if it is valid
        for all relational valuations.
        \item A predicate is \keyword{expressible in \(\allRELs\)} if there
        exists a pair of KAT terms that express the predicate in all
        relational KATs.
    \end{itemize}
\end{definition}

One of the most important results in~\cite{goos_kleene_1997} is the completeness 
of KAT over relational KAT:
\[\validUnderModels{\allRELs}{\synt{t_1} = \synt{t_2}} \iff
  \validUnderModels{\allKATs}{\synt{t_1} = \synt{t_2}}\] This means all
equalities that are valid in all relational KAT can be deduced using just the
theory of KAT\@.  This result relies crucially on the construction of so-called
\emph{language KATs}, whose carrier sets are guarded terms of actions.
\citet{goos_kleene_1997} showed that every KAT term can be interpreted in such
KATs~\cite[Section~3]{goos_kleene_1997} and, following from~\citet{Pratt_1980},
proved the completeness of relational KAT by the existence of an injective
homomorphism from any language KAT to a relational KAT and the completeness of
language KATs\@.  We will use similar techniques to obtain the completeness
results in \Cref{sec: properties of TopKAT}.

\subsection{Hoare Logic}\label{sec:incorrectness-and-hoare}

Hoare logic is a fundamental tool for specifying and proving the correctness of
while-like programs. Following \citet{kozen_hoare_2000}, we consider here
\emph{propositional Hoare logic}, which involves partial correctness Hoare
triples \(\hoareTriple{b}{p}{c}\) consisting of atomic propositions,
propositional connectives and while-like programs.  As usual, a Hoare triple
\(\hoareTriple{b}{p}{c}\) means that if the program \(p\) terminates when run on
a memory satisfying \(b\), it will result in a memory satisfying \(c\).
\Cref{fig:HL} shows the rules of propositional Hoare logic, which differ from
the classical setting in their omission of the assignment rule.

\begin{figure}
    \begin{mathpar}
        \inferrule[Composition]
        {\hoareTriple{a}{p}{b} \\ \hoareTriple{b}{q}{c}}
        {\hoareTriple{a}{p;q}{c}}
    
        \and 
        \inferrule[Conditional]
        {\hoareTriple{b\land c}{p}{d} \\ \hoareTriple{\neg b\land c}{q}{d}}
        {\hoareTriple{c}{\comITE{b}{p}{q}}{d}}

        \and
        \inferrule[While]
        {\hoareTriple{b\land c}{p}{c}}
        {\hoareTriple{c}{\comWhile{b}{p}}{\neg b\land c}}

        \and
        \inferrule[Consequence]
        {b' \limplies b \\ \hoareTriple{b}{p}{c} \\ c \limplies c'}
        {\hoareTriple{b'}{p}{c'}}

      \end{mathpar}
    \caption{Propositional Hoare logic}\label{fig:HL}
\end{figure}

In its essence, Hoare logic is an \emph{over-approximation} logic.  To see this,
it is convenient to think about a program \(p\) as a relation between input
memories and output memories, and to think about predicates \(b\) and \(c\) as
sets of states.  Given a program \(p\) and a predicate \(b\), we can write
\(\post(p)(b)\) for the set of post-states, that is
\[\post(p)(b)=\{ x \mid \exists y\in b, (y,x)\in p \}.\]
A partial-correctness Hoare triple \(\hoareTriple{b}{p}{c}\) is valid
iff \[\post(p)(b) \subseteq c.\] In words, \(c\) over-approximates the set of
memories which can be obtained from \(b\) by running the program \(p\).  This
condition can be expressed by means of the codomain of a relation: if we set
\[\codomain{r} \definedAs \{y \mid \exists x \in X, (x, y) \in r\},\]
then, for all a relational KAT \((\algebra{R}, \algebra{b})\), 
\(b, c \in \algebra{B}\) and \(p \in \algebra{R}\),
\[\hoareTriple{b}{p}{c} \definedAs \codomain{b p} \subseteq \codomain{c}.\]

\citet{kozen_hoare_2000} showed that we can reason about the 
partial correctness of propositional Hoare logic in KAT\@. 
To do this  we can use tests to represent pre and post-conditions,
thus encode a partial correctness propositional Hoare triple
\(\hoareTriple{b}{p}{c}\) as the KAT equality:
\[b p \compl{c} = 0\]
or equivalently \[b p = b p c.\]

Indeed, we can show that in all relational KATs \((\algebra{R}, \algebra{B})\) and
\(p \in \algebra{R}, b, c \in \algebra{B}\),
\[\hoareTriple{b}{p}{c} \iff bp = bpc \iff b p \compl{c} = 0.\]
Given that KAT is complete over relational KAT, we can determine the relational
validity of some propositional Hoare triples by the equational theory of KAT\@.

\subsection{Incorrectness Logic}
%
\citet{ohearn_incorrectness_2020} proposed incorrectness logic to reason about
incorrect programs.  This logic is also related to earlier works by
\citet{barthe_reverse_2011}.  Like Hoare logic, incorrectness logic is built on
triples of the form \(\incorTriple{b}{p}{c}\), which denote \emph{incorrectness
  specifications}, where \(c\) is a set of undesirable final states and \(b\) is a
precondition. Intuitively, such a triple says that every memory in \(c\) needs
to be reachable from \(b\) by executing \(p\).

\begin{figure}
    \begin{mathpar}
        \inferrule[Empty]
        {\\}{\incorTriple{b}{p}{\epsilon: \falsehood}}

        \and
        \inferrule[Consequence]
        {b \limplies b' \\ \incorTriple{b}{p}{\epsilon: c} \\ c' \limplies c}
        {\incorTriple{b'}{p}{\epsilon: c'}}

        \and
        \inferrule[Disjunction]
        {\incorTriple{b_1}{p}{\epsilon: c_1} \\ \incorTriple{b_2}{p}{\epsilon: c_2}}
        {\incorTriple{b_1 \lor b_2}{p}{\epsilon: (c_1 \lor c_2)}}

        \and
        \inferrule[Identity]
        {\\} {\incorTriple{b}{\comSkip}{\okState: b, \errState: 0}}

        \and
        \inferrule[Composition-Fail]
        {\incorTriple{a}{p}{\errState: b}}
        {\incorTriple{a}{p;q}{\errState: b}}

        \and
        \inferrule[Composition-Normal]
        {\incorTriple{a}{p}{\okState: b} \\ \incorTriple{b}{p}{\epsilon: c}}
        {\incorTriple{a}{p;q}{\epsilon: c}}

        \and
        \inferrule[Choice-Left]
        {\incorTriple{b}{p}{\epsilon: c}}
        {\incorTriple{b}{p + q}{\epsilon: c}}

        \and
        \inferrule[Choice-Right]
        {\incorTriple{b}{q}{\epsilon: c}}
        {\incorTriple{b}{p + q}{\epsilon: c}}

        \and
        \inferrule[Assume]
        {\\}
        {\incorTriple{a}{\comAssume{b}}{\okState: a \land b, \errState: 0}}

        \and
        \inferrule[Error]
        {\\}{\incorTriple{b}{\fail}{\errState: b}}

        \and
        \inferrule[Iter-Zero]
        {\\}{\incorTriple{b}{\starOf{p}}{\okState: b}}

        \and
        \inferrule[Iter-NonZero]
        {\incorTriple{b}{\starOf{p};p}{\epsilon: c}}
        {\incorTriple{b}{\starOf{p}}{\epsilon: c}}

        \and
        \inferrule[Iter-Dependent]
        {\forall n \in \nat,~ \incorTriple{b(n)}{p}{\okState: b(n+1)}}
        {\incorTriple{b(0)}{\starOf{p}}{\okState: \exists n, b(n)}}
    \end{mathpar}
    \caption{Generic incorrectness logic proof rules~\cite{ohearn_incorrectness_2020}}\label{fig: orignal proof rules of inc logic}
\end{figure}

If Hoare logic is an ``over-approximation'' logic, incorrectness logic is an
``under-approximation'' logic: an incorrectness triple \(\incorTriple{b}{p}{c}\)
is valid if and only if
\[\post(p)(b) \supseteq c.\]

In other words, the post-condition \(c\) ``under-approximates'' the canonical
post condition of \(b\) after executing \(p\).
When we only look at program that terminates normally, 
the previous definition can be expressed in relational KAT as 
\[\codomain{bp} \supseteq \codomain{c}.\]
As mentioned by \citet[Section~2]{ohearn_incorrectness_2020},
the definitions of incorrectness and Hoare triples are highly symmetric:
\begin{align*}
    \hoareTriple{b}{p}{c} & \definedAs \codomain{b p} \subseteq \codomain{c} \\
    \incorTriple{b}{p}{c} & \definedAs \codomain{b p} \supseteq \codomain{c}
\end{align*}

Since incorrect programs often lead to explicit errors,
\citet{ohearn_incorrectness_2020} also considered incorrectness triples
\(\incorTriple{b}{p}{\errState: c}\), which mean that, in addition to satisfying
$c$, we require that the final states of $p$ lead to an error.  More generally,
we use the notation $\incorTriple{b}{p}{\epsilon: c}$, where the \emph{error
  code} \(\epsilon \in \{\okState, \errState\}\) signals whether the program
terminated normally or not; hence, the unqualified notation
\(\incorTriple{b}{p}{c}\) is simply a shorthand for
\(\incorTriple{b}{p}{\okState: c}\), when abnormal termination is not a
concern. Informally, such general triples mean that
\[\incorTriple{b}{p}{\epsilon: c} \definedAs \codomain{b p} \supseteq \codomain{(\epsilon: c)}.\]
In \Cref{sec: failkat and error}, we will give a more concrete definition of
this semantics in an extension of KAT\@.

The generic proof rules of incorrectness logic are listed in \Cref{fig: orignal
  proof rules of inc logic}.
Following \citet{ohearn_incorrectness_2020}, we formulate
incorrectness logic for a language of commands which is essentially
the same of KAT terms. Conditionals and loops can be encoded with an
encoding that is similar to the one given by
\citet{kozen_kleene_1997}.
\begin{align*}
    \comITE{b}{p}{q} & \definedAs  (\comAssume{b};p) + (\comAssume{\compl{b}}; q) \\
    \comWhile{b}{p} & \definedAs \starOf{(\comAssume{b}; p)} ;\comAssume{\compl{b}}
\end{align*}
Following \citet{ohearn_incorrectness_2020}
we also use
\(\incorTriple{b}{p}{\okState: c_1, \errState: c_2}\) as a shorthand for two
different rules. For example, the rule
\begin{mathpar}
    \inferrule[Unit]
    {\\}{\incorTriple{a}{1}{\okState: a, \errState: 0}}
\end{mathpar}
stands for the two rules
\begin{mathpar}
    \inferrule[Unit-Ok]
    {\\}{\incorTriple{a}{1}{\okState: a}}

    \inferrule[Unit-Er]
    {\\}{\incorTriple{a}{1}{\errState: 0}}.
  \end{mathpar}
The fragment of incorrectness logic we consider here is \emph{propositional} in the sense of \citet{kozen_hoare_2000}. In particular, this presentation omits  rules for variables and mutation~\cite{ohearn_incorrectness_2020}. 

\section{Formulating Incorrectness Logic}\label{sec: formulating incorrectness logic}

We might hope that the symmetry between Hoare logic and incorrectness logic
would help us express incorrectness triples by adapting the formulation of
\citet{kozen_hoare_2000}. However,
it is not obvious how we can exploit this symmetry, since it involves the
codomain operation, which does not appear in the formulation of
\citet{kozen_hoare_2000}.  
This difficulty, unfortunately, is fundamental: KAT cannot express
incorrectness logic.

\begin{definition}\label{def: relational validity of incorrectness triple}
  Given a relational KAT \((\algebra{R}, \algebra{B})\), \(p \in \algebra{R}\)
  and \(b, c \in \algebra{B}\), an Incorrectness Triple
  \(\incorTriple{b}{p}{c}\) is \keyword{valid} with respect to
  \((\algebra{R}, \algebra{B})\), denoted
  \((\algebra{R}, \algebra{B}) \models \incorTriple{b}{p}{c}\), if
  \[\codomain{bp} \supseteq \codomain{c}\]
  we consider the predicate of incorrectness triple:
  \(inc(b, p, c) \definedAs \incorTriple{b}{p}{c}\).  We write \(inc\) over
  primitive tests \(\synt{b}, \synt{c}\) and primitive action \(\synt{p}\) as
  \(\incorTriple{\synt{b}}{\synt{p}}{\synt{c}}\).
\end{definition}

To show that incorrectness triple cannot be formulated using equality of KAT
terms, we only need to show that \(\incorTriple{\synt{b}}{\synt{p}}{\synt{c}}\)
cannot be expressed in \(\allRELs\).  More explicitly, we need to show that for
all \(K, B\) where \(\synt{p} \in K\) and \(\synt{b}, \synt{c} \in B\), there
does \emph{not} exist a pair of terms
\(\synt{t_1}, \synt{t_2} \in \KATTerm{K}{B}\) s.t.\ for all relational
valuations \(u\):
\[\validUnderVal{u}{\synt{t_1} = \synt{t_2}} \iff \validUnderVal{u}{\incorTriple{\synt{b}}{\synt{p}}{\synt{c}}}\]

\begin{theoremEnd}[normal]{theorem}\label{the: KAT not able to express incorrectness logic}
    \(\incorTriple{\synt{b}}{\synt{p}}{\synt{c}}\)
    cannot be expressed in \(\allRELs\).
\end{theoremEnd}

\begin{figure}
    \centering
    \begin{tikzcd}[row sep=1.5em, column sep=1.5em]
        (x, x) \in \interpWith{u}{\synt{t}_1}
            \arrow[rrr, "\interpWith{u}{\synt{t}_1} = \interpWith{u}{\synt{t}_2}"] 
            &&& 
        (x, x) \in \interpWith{u}{\synt{t}_2} \\
        & \interpWith{u}{\synt{t}_1} 
            \arrow[r, equal]
        & \interpWith{u}{\synt{t}_2} & \\
        & \interpWith{u_{\emptyset}}{\synt{t}_1} 
            \arrow[u, hook, "\text{monotonicity}"]
            \arrow[r, equal, "/" marking]
        & \interpWith{u_{\emptyset}}{\synt{t}_2} 
            \arrow[u, hook, "\text{monotonicity}"']
        & \\
        (x, x) \in \interpWith{u_{\emptyset}}{\synt{t}_1} 
            \arrow[uuu, "\text{by monotonicity}"] &&& 
        (x, x) \not \in \interpWith{u_{\emptyset}}{\synt{t}_2} 
            \arrow[uuu, dash, "\text{contradiction}"', "\times"marking]\\
    \end{tikzcd}
    \caption{Relation of interpretations \(u\) and \(u_{\emptyset}\)}\label{fig: proof of KAT cannot encode IL}
\end{figure}
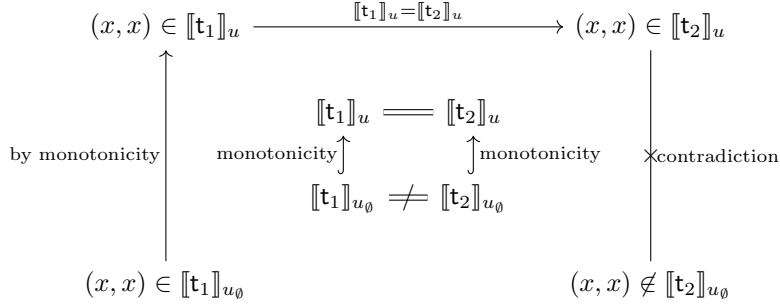

\begin{proofEnd}
\iffull
    First, by \Cref{the: redundancy of alphabet},
    we only need to show that there does not exist
    \[\synt{t_1}, {\synt{t_2}} \in \KATTerm{\{\synt{p}\}}{\{\synt{b}, \synt{c}\}}\]
    that can express incorrectness logic.

    Let's assume that there exist \({\synt{t_1}}\) and \({\synt{t_2}}\) 
    in \(\KATTerm{\{\synt{p}\}}{\{\synt{b}, \synt{c}\}}\)
    such that
    \[\validUnderVal{u}{{\synt{t_1}} = {\synt{t_2}}}
    \iff \validUnderVal{u}{\incorTriple{\synt{b}}{\synt{p}}{\synt{c}}}\] holds for
    all relational valuations \(u\).

    Consider the relational KAT \(\algebra{R}\) 
    that contains all the relations and predicates over \(\{0, 1\}\).
    We will construct a pair of valuations on \(\synt{b}, \synt{c}, \synt{p}\) 
    to show a contradiction: 
    \begin{align*}
        u_{\emptyset}(\synt{p}) & \definedAs \emptyset 
        & u(\synt{p}) & \definedAs \{(0, 1)\} \\
        u_{\emptyset}(\synt{b}) & \definedAs \{(0, 0)\}
        & u(\synt{b}) & \definedAs \{(0, 0)\} \\
        u_{\emptyset}(\synt{c}) & \definedAs \{(1, 1)\}
        & u(\synt{c}) & \definedAs \{(1, 1)\}.
    \end{align*}
    Where the incorrectness triple
    \(\incorTriple{\synt{b}}{\synt{p}}{\synt{c}}\) is valid with valuation
    \(u\), but not with \(u_{\emptyset}\).  And the only difference between
    \(u\) and \(u_{\emptyset}\) is that \(\synt{p}\) is mapped to \(\emptyset\)
    in \(u_{\emptyset}\).

    Since \(u_{\emptyset}\) valuates the only action variable \(\synt{p}\) as \(\emptyset\),
    By \Cref{the: only one action var} 
    all the elements of \(\interpWith{u_{\emptyset}}{\synt{t}_1}\) must be of the form \((x, x)\).
    Then because the incorrectness triple \(\incorTriple{\synt{b}}{\synt{p}}{\synt{c}}\)
    is invalid with \(u_{\emptyset}\),
    \[\interpWith{u_{\emptyset}}{\synt{t}_1} \neq \interpWith{u_{\emptyset}}{\synt{t}_2}.\]
    Without loss of generality, assume that
    \((x, x) \in \interpWith{u_{\emptyset}}{\synt{t}_1}\), but not in
    \(\interpWith{u_{\emptyset}}{\synt{t}_2}\).  Then by monotonicity of
    interpretation (\Cref{the: monotonicity of relational interpretation}),
    \((x, x) \in \interpWith{u}{\synt{t}_1}\).  We will derive a contradiction
    from the fact that \((x, x) \in \interpWith{u}{\synt{t}_2}\) but
    \((x, x) \in \interpWith{u_{\emptyset}}{\synt{t}_2}\).  We can summarize the
    above strategy using \Cref{fig: proof of KAT cannot encode IL}.

    Because \((x, x) \not \in \interpWith{u_{\emptyset}}{\synt{t_2}}\)
    and the only element of the action \(u(\synt{p})\) is \((1, 0)\),
    none of the conditions in \Cref{the: KAT inccorrectness logic core lemma} is satisfied,
    therefore \((x, x)\) cannot be in \(\interpWith{u}{\synt{t_2}}\),
    which contradicts the earlier result stating \((x, x) \in \interpWith{u}{\synt{t_2}}\).
\fi
\end{proofEnd}

One strategy for exploiting the symmetry between Hoare and incorrectness logic
is to extend KAT with a codomain operator.  Similar ideas have been explored in
prior
work~\cite{fitzgerald_modal_2016,fahrenberg_domain_2021,desharnais_modal_2004}. However,
rather than adopting a full-fledged codomain operator, it suffices for our
purposes to consider a equational theory that only extends KAT with a top
element.  Such an algebraic structure has also been considered in prior
work~\cite[Section~4]{esparza_equational_2017}, though for a different purpose.

\begin{definition}[KAT with a Top Element]
  A KAT with top, or TopKAT, is a KAT \(\algebra{K}\) that contains a largest
  element \(\top\); that is, for all elements \(p \in \algebra{K}\),
  \[\top \geq p.\]
  A relational TopKAT over \(X\) is a relational KAT 
  where the top element is the complete relation \(X \times X\).
\end{definition}

\begin{theoremEnd}[normal]{theorem}[TopKAT can Express Codomain]\label{the: top element can express domain}
    For all relational TopKATs \(\algebra{R}\),
    and \(p, q \in \algebra{R}\) The following is true:
    \[\top p = \top q \iff \codomain{p} = \codomain{q}\]
    and 
    \[\top p \leq \top q \iff \codomain{p} \subseteq \codomain{q}.\]
\end{theoremEnd}

\begin{proofEnd}
  For all relations \(r \subseteq X \times X\):
    \begin{align*}
        \top \thenCompo r 
        & = \{(z, y) \mid z \in X \land \exists x, (x, y) \in r\} 
        = \{(z, y) \mid z \in X \land y \in \codomain{r}\} 
    \end{align*}
    Therefore for two relation \(p, q\), we will have 
    \begin{align*}
        \top \thenCompo q = \top \thenCompo p & \iff 
        \{(z, y) \mid z \in X \land y \in \codomain{q}\}
        = \{(z, y) \mid z \in X \land y \in \codomain{p}\} \\
        & \iff \codomain{q} = \codomain{p} \\
        \top \thenCompo q \leq \top \thenCompo p & \iff 
        \{(z, y) \mid z \in X \land y \in \codomain{q}\}
        \subseteq \{(z, y) \mid z \in X \land y \in \codomain{p}\} \\
        & \iff \codomain{q} \subseteq \codomain{p} 
    \end{align*}
\end{proofEnd}

\begin{theoremEnd}[normal]{corollary}\label{the: topkat can express incorrectness logic normal termination}
    For all relational TopKATs \((\algebra{R}, \algebra{B})\),
    and \(p \in \algebra{R}\), \(b, c \in \algebra{B}\)
    we have the following:
    \[{\top bp \geq \top c} \iff {\incorTriple{b}{p}{c}}\]
\end{theoremEnd}

Notice that the left hand side of this equivalence makes sense in all TopKATs,
not just relational ones.
Thus it is natural to generalize the definition of incorrectness triple to all TopKATs
using the inequality \(\top bp \geq \top c\).
As a matter of fact, just by using the equational theory of TopKAT,
we can find several equivalent formulations of incorrectness triple:

\begin{theoremEnd}{theorem}[Equivalent Formulation of Incorrectness Logic]
  Given a TopKAT \((\algebra{K}, \algebra{B})\), where \(p \in \algebra{K}\) and
  \(b, c \in \algebra{B}\), we say that an incorrectness triple
  \(\incorTriple{b}{p}{c}\) is valid if the following equivalent conditions are
  met:
  \[\top b p \geq \top c \iff \top b p \geq c \iff \top b p c = \top c\]
\end{theoremEnd}

\begin{proofEnd}
We show that these conditions are equivalent in all TopKATs.
\begin{itemize}
    \item First show \[\top b p \geq \top c \iff \top b p \geq c,\]
    the \(\implies\) is true because \(\top \geq 1\):
    \[\top b p \geq \top c \geq c,\]
    the \(\impliedby\) can be proven by idempotency of \(\top\):
    \[\top b p = \top \top b p \geq \top c.\]
    \item Then show \[\top b p \geq \top c \iff \top b p c = \top c,\]
    the \(\impliedby\) can be shown by \(1 \geq c\):
    \[\top b p \geq \top b p c = \top c,\]
    the \(\implies\) is by two inequalities:
    \begin{align*}
        \top b p \leq \top & \implies \top b p c \leq \top c \\
        c \leq 1 \land \top b p \geq \top b p c & \implies \top b p \geq \top b p c = \top c.\\
    \end{align*}
    Therefore we have our conclusion \(\top b p c = \top c\).
\end{itemize}
\end{proofEnd}

A similar encoding involving \(\top\) was also mentioned by \citet[Section~5.3]{ohearn_incorrectness_2020}.

We want to show that this abstraction is enough to capture all the rules of
incorrectness logic.  Indeed most of the the rules are formulated using program
operations expressible in \(\allKATs\) ~\cite{ohearn_incorrectness_2020}.  We
focus here on the proof rules concerning normal program termination, and we will
further investigate the rules with error in \Cref{sec: failkat and error}.

\begin{figure}
    \begin{mathpar}
        \inferrule[Empty]
        {\\}{\incorTriple{b}{p}{0}}
    
        \and
        \inferrule[Consequence]
        {b \leq b' \\ \incorTriple{b}{p}{c} \\ c' \leq c}
        {\incorTriple{b'}{p}{c'}}
    
        \and 
        \inferrule[Disjunction]
        {\incorTriple{b_1}{p}{c_1} \\ \incorTriple{b_2}{p}{c_2}}
        {\incorTriple{b_1 + b_2}{p}{c_1 + c_2}}
    
        \and 
        \inferrule[Identity]
        {\\} {\incorTriple{b}{1}{b}}
    
        \and 
        \inferrule[Composition]
        {\incorTriple{a}{p}{b} \\ \incorTriple{b}{q}{c}}
        {\incorTriple{a}{pq}{c}}
    
        \and 
        \inferrule[Choice-Left]
        {\incorTriple{a}{p}{b} }
        {\incorTriple{a}{p + q}{b}}
    
        \and 
        \inferrule[Choice-Right]
        {\incorTriple{a}{q}{b} }
        {\incorTriple{a}{p + q}{b}}
    
        \and 
        \inferrule[Assume]
        {\\}
        {\incorTriple{b}{c}{bc}}
    
        \and 
        \inferrule[Iter-Zero]
        {\\}{\incorTriple{b}{\comIter{p}}{b}}
    
        \and 
        \inferrule[Iter-NonZero]
        {\incorTriple{b}{\comIter{p}p}{c}}{\incorTriple{b}{\comIter{p}}{c}}
    
        \and
        \inferrule[Iter-Dependent]
        {\forall n \in \nat, \incorTriple{b_n}{p}{b_{n + 1}}}
        {\incorTriple{b_0}{\comIter{p}}{\sup_{n \in \nat} b_{n}}}
    \end{mathpar}
    \caption{Incorrectness logic with normal termination}\label{fig: normal terminating Inc rule}
\end{figure}

In \Cref{fig: normal terminating Inc rule}, we present all the rules of
incorrectness logic with normal termination using the syntax of KAT.  (Note
that this differs slightly from \citeauthor{ohearn_incorrectness_2020}'s
original notation; for instance, the sequencing operator \(p;q\) corresponds to
multiplication \(pq\) in KAT, and the logical implication \(b \limplies b'\)
corresponds to order \(b \leq b'\).)  For the Iter-dependent rule (Backwards
Variant in~\cite{ohearn_incorrectness_2020}), the function \(b(n)\) corresponds
to a sequence of tests \((b_n)_{n \in \nat}\), and the existential
\(\exists n, b(n)\) corresponds to the infinite disjunction of all these
\(b_n\), which we express as \(\sup_{n \in \nat} b_n\).  (Note that \(\sup\)
does not exist in KATs, TopKATs, or boolean algebras in general, and this rule
implicitly assumes that \(\sup_{n \in \nat} b(n)\) exists.)

\begin{theoremEnd}[normal]{theorem}[Soundness of \(\okState\) State Rules]\label{the: incor logic rule sound ok}
    All the rules of \Cref{fig: normal terminating Inc rule}
    are derivable in all TopKATs.
\end{theoremEnd}

\begin{proofEnd}
\iffull
    The proofs of each individual rules as follows:
    \begin{itemize}
        \item \itemTitle{Empty Rule:}
            because 0 is the smallest element 
        \item \itemTitle{Consequence Rule:}
            because multiplication preserves order
            and \(b' \geq b\), therefore \(\top b' p \geq \top b p\).
            Therefore 
            \[\top b' p \geq \top b p \geq c \geq c'\] 
        \item \itemTitle{Disjunction Rule:}
            because addition preserves order, therefore 
            \[\top b_1 p + \top b_2 p \geq c_1 + c_2\]
            and by distributivity, we have 
            \[\top (b_1 + b_2) p \geq c_1 + c_2\]
        \item \itemTitle{Identity Rule:}
            because \(\top \geq 1\),
            therefore \[\top b 1 \geq 1 b 1 \geq b\]
        \item \itemTitle{Composition Rule:}
            first, by idempotency of \(\top\)
            and associativity of multiplication, we have 
            \[\top a (p q) = \top \top a p q\]
            By order preserving of multiplication, 
            and \(\top a p \geq b\),
            we have \[\top a (p q) = \top \top a p q \geq \top b q \geq c\]
        \item \itemTitle{Choice Left and Choice Right Rule:}
            by distributivity, we have 
            \[\top b (p + q) 
            = \top b p + \top b q 
            \geq \top b p \geq c\]
            and 
            \[\top b (p + q) 
            = \top b p + \top b q 
            \geq \top b q \geq c\]
        \item \itemTitle{Assume Rule:}
            since \(\top \geq 1\), 
            we have \[\top b c = \top b c \geq 1 b c = bc\]
        \item \itemTitle{Iter-Zero Rule:}
            since \(\starOf{p} = p \starOf{p} + 1\),
            we have 
            \[\top b \starOf{p} 
            = \top b (p \starOf{p} + 1)
            \geq \top b 1 \geq b\]
        \item \itemTitle{Iter-NonZero Rule:}
            since \(\starOf{p} = p \starOf{p} + 1\),
            we have 
            \[\top b \starOf{p} 
            = \top b (p \starOf{p} + 1)
            \geq \top b p \starOf{p} \geq c\]
        \item \itemTitle{Iter-Dependent Rule:}
            by definition of \(\sup\),
            in order to show \(\top b_{0} \starOf{p} \geq \sup_{n \in \nat} b_{n}\)
            all we need to show is that for all \(m \in \nat\)
            \(\top b_{0} \starOf{p} \geq b_{m}\).
            We prove this by induction on \(m\).
            \begin{itemize}
                \item \itemTitle{Base Case:}
                    we need to show \(\top b_{0} \starOf{p} \geq b_{0}\).
                    Because \(\starOf{p} = 1 + p\starOf{p}\),
                    therefore 
                    \[\top b_{0} \starOf{p} 
                    = \top b_{0} (1 + p\starOf{p})
                    \geq \top b_{0} 1 \geq 1 b_{0} 1 = b_{0}\]
                \item \itemTitle{Inductive Case:}
                    if \(\top b_{0} \starOf{p} \geq b_{n}\),
                    we need to show \(\top b_{0} \starOf{p} \geq b_{n+1}\).
                    By premise we have \(\top b_{n} p \geq b_{n + 1}\)
                    Therefore by idempotency of the top element,
                    we have 
                    \begin{align*}
                        \top b_{0} \starOf{p} 
                        & = \top \top b_{0} (1 + \starOf{p} p) \\
                        & \geq \top \top b_{0} \starOf{p} p
                        & 1 + \starOf{p} p \geq \starOf{p} \\
                        & \geq \top b_{n} p
                        & \text{by induction hypothesis} \\
                        & \geq b_{n+1}
                        & \text{by premise}
                    \end{align*}
            \end{itemize}   
    \end{itemize}
\fi
\end{proofEnd}

\paragraph*{An alternative encoding for Hoare logic}

Since \(\allTopKATs\) can express codomain (\Cref{the: top element can express
  domain}), we can exploit the symmetry between incorrectness logic and Hoare
logic (\Cref{sec:incorrectness-and-hoare}) to give another encoding of Hoare
logic in \(\allTopKATs\): \[\top b p \leq \top c.\] This is equivalent to the
original encoding $bp = bpc$ proposed for KATs~\cite{kozen_hoare_2000} in \emph{all TopKAT},
not just relational ones.  
Since the proof rules of propositional Hoare logic are sound for that
encoding~\cite[Theorem~3.1]{kozen_hoare_2000}, they are also sound for ours.

\begin{theoremEnd}[normal]{theorem}[Equivalence of Hoare Logic Formulations]\label{the: hoare logic formulation equivalent}
  For all TopKATs \((\algebra{K}, \algebra{B})\), and three elements
  \(p \in \algebra{K}\) and \(b, c \in \algebra{B}\), the following inequalities
  are equivalent:
    \[b p \leq \top c  \iff  \top b p \leq \top c \iff b p = b p c.\]
\end{theoremEnd}

\begin{proofEnd}
\begin{itemize}
    \item We first show the equivalence:
        \[b p \leq \top c  \iff  \top b p \leq \top c.\]
        \(\impliedby\) is by \(1 \leq \top\),
        thus \[b p \leq \top b p \leq \top c,\]
        and \(\implies\) is by the idempotency of \(\top\),
        multiply both side with \(\top\), we have 
        \[\top b p \leq \top \top c = \top c.\]
    \item Then we show the equivalence
        \[b p \leq \top c  \iff  b p = b p c.\]
        Start with \(\impliedby\), because \(\top \geq b p\), we have
        \[b p = b p c \leq \top c.\]
        Then \(\implies\), we start from \(b p \leq \top c\),
        we first multiply \(\compl{c}\) on both side,
        \[b p \compl{c} \leq \top c \compl{c} = \top 0 = 0,\]
        and because \(b p \compl{c} \geq 0\) (\(0\) is the smallest element),
        we have \[b p \compl{c} = 0.\]
        Finally, add \(b p c\) to both side, we have
        \[b p = b p (\compl{c} + c) = 0 + b p c = b p c.\]
\end{itemize}
\end{proofEnd}

\section{Properties of TopKATs}\label{sec: properties of TopKAT}

In the previous section, we have shown that the theory of TopKAT subsumes incorrectness logic. In this section, we study some of the properties of its algebraic equational theory.
First, we will show that TopKAT is \emph{incomplete} with respect to relational
TopKATs: there are identities that are valid for every relational TopKAT that
cannot be proven using the TopKAT identities alone. This contrasts with what
happens for KAT, which is complete for relational KATs.
We will then show the completeness of TopKATs with respect to a class of
\emph{language-based} TopKATs and also with respect to a more general class of
\emph{relational} TopKATs, as well as the decidability of TopKAT equalities.
Finally we will introduce the concept of equational expressiveness,
and show that the general relational TopKAT has the same expressiveness as relational KAT,
hence cannot express incorrectness logic.

\subsection{Incompleteness with Respect to Relational TopKATs}
We can naturally extend the definition of term,
and primitives of a term from KAT to TopKAT\@.
An \keyword{alphabet} \((K, B)\) is two disjoint sets \(K\) and \(B\), 
where the elements of \(K\) are called \keyword{primitive actions}, 
and the elements of \(B\) are called \keyword{primitive tests}. 
The set \(\TKATTerm{K}{B}\) of \keyword{TopKAT terms} over
an alphabet \(K, B\) is generated by the following grammar:
\[\synt{t} \definedAs
\synt{p} \in K \mid 
\synt{b} \in B \mid 
\top \mid 0 \mid 1 \mid
\synt{t_1} + \synt{t_2} \mid
\synt{t_1} \synt{t_2} \mid 
\starOf{\synt{t}} \mid 
\compl{\synt{t_b}},\]
where \(\synt{t_b}\) does not contain primitive actions.

Similarly we can extend the notions of valuation and interpretation.  Given an
alphabet \(K, B\), and a TopKAT \(\algebra{K}\), a valuation for TopKAT terms is
a function \(u: K \union B \to \algebra{K}\).  The interpretation
\(\interpWith{u}{-}: \TKATTerm{K}{B} \to \algebra{K}\) is generated by \(u\), as
in \Cref{def:kat-valuation}.

We start with our negative result, which is the incompleteness over relational TopKATs.

\begin{theoremEnd}[normal]{theorem}\label{the: topkatstar incomplete over relational model}
  \(\allTopKATs\) is incomplete over \(\allTopRELs\): the formulas
  \[(\top \synt{p}) (\top \synt{p}) = \top \synt{p} \tand 
  \synt{p} \top \synt{p} \geq \synt{p}\] 
  are valid in every relational TopKAT, but not in every TopKAT\@.
\end{theoremEnd}

\begin{proof}
  We can show that the \((\top \synt{p}) (\top \synt{p}) = \top \synt{p}\) 
  holds in every relational TopKAT simply by
  unfolding the definitions.  To exhibit a TopKAT where it does not hold,
  consider the following counterexample.  We take a relational KAT over
  \(\{0,1\}\) whose largest element is
  \begin{align*}
    \top & \definedAs \{(0,0), (1,1), (0,1)\}.
  \end{align*}
  (Note that this KAT is \emph{not} a relational TopKAT\@: the largest
  element is not the complete relation.)
  Then, by taking a valuation \(u(\synt{p}) \definedAs\{(0, 1)\}\), we see that
  \begin{align*}
    \interpWith{u}{\top \synt{p}} & = \{(0, 1)\} \\
    \interpWith{u}{(\top \synt{p}) (\top \synt{p})} & 
        = \{(0, 1)\} \thenCompo \{(0, 1)\} 
        = \emptyset \neq \interpWith{u}{\top \synt{p}}
  \end{align*}

  And the same counter example also works for 
  \(\synt{p} \top \synt{p} \geq \synt{p}\).
\end{proof}

The incompleteness result might be discouraging, since we might not be able to
derive some valid theorems of incorrectness logic using TopKAT. However in
\Cref{the: incor logic rule sound ok} and \Cref{the: incorrectness logic fail
  rule sound}, we show that all the rules of incorrectness logic are
derivable using either the equational theory of TopKAT or FailTopKAT\@.  Thus,
our framework is at least as strong as the structural proof rules presented
by~\citet[Figure~2]{ohearn_incorrectness_2020}.

\subsection{Completeness and Decidability of \allTopKATs}

Language models are essential in various extensions of Kleene algebra, since they
are the basis of many completeness and decidability
proofs~\cite{goos_kleene_1997,hutchison_kleene_2014}.  In this section, we adapt
this idea to the setting of TopKATs. The construction 
follows~\citet[Section~3]{goos_kleene_1997}, except the \(\top\) element is 
treated as a primitive action in the language model.

\begin{definition}
    For an alphabet \(K, B\), where
    \(B = \{\synt{b_0}, \synt{b_1}, \dots \synt{b_n}\}\) a \keyword{minimal test}
    (a smallest non-zero test in the free TopKAT, which is called ``atom''
    by~\citet{goos_kleene_1997}), is a term of the following form:
    \[\hat{\synt{b_0}} \hat{\synt{b_1}} \dots \hat{\synt{b_n}} \twhere 
    \forall i \in \{0, 1, \dots, n\}, \hat{\synt{b_i}} \in \{\synt{b_i},
    \compl{\synt{b_i}}\}\] We let \(\alpha, \beta, \gamma\) range over minimal
    tests.  We will write \(\unitTopLangModel\) for the set of all minimal tests over an
    alphabet \(K, B\), when \(K, B\) can be inferred from context. 
    When the boolean alphabet is empty, 
    then \(\unitTopLangModel\) will only contain the empty product, which is 1.
\end{definition}

\begin{definition}
    For an alphabet \(K, B\) a \keyword{guarded term} is a term of the following form:
    \[\alpha_0 \synt{p_1} \alpha_1 \dots \synt{p_n} \alpha_n\]
    where \(\synt{p_1}, \synt{p_2}, \dots \synt{p_n} \in K \union \{\top\}\) and 
    \(\alpha_1, \alpha_2, \dots \alpha_n \in \unitTopLangModel\).

    We call the set of all guarded terms \(\topGuardedTerm\),
    as this definition includes \(\top\) as a primitive action.
    We sometimes write a guarded term as \(\synt{s} \alpha\) or \(\alpha \synt{s}\), 
    to represent the concatenation of the term \(\synt{s}\) with a minimal test \(\alpha\).
\end{definition}

\begin{definition}
  The \keyword{coalesced product} is a partial binary operation on
  \(\topGuardedTerm\) defined as follows:
    \[
        \synt{s_1} \diamond \synt{s_2} = 
        \begin{cases}
        \hat{\synt{s_1}} \alpha \hat{\synt{s_2}} & 
            \synt{s_1} = \hat{\synt{s_1}} \alpha \tand \synt{s_2} = \alpha \hat{\synt{s_2}} \\
        \text{undefined} & \text{otherwise}
        \end{cases}
    \]  
    This definition can naturally extend to subsets of \(\topGuardedTerm\) as a
    total binary operation on subsets of \(\topGuardedTerm\):
    \[S_1 \diamond S_2 = 
    \{\synt{s_1} \diamond \synt{s_2} \mid 
        \synt{s_1} \in S_1, \synt{s_2} \in S_2, \synt{s_1} \diamond \synt{s_2} \text{ is defined}\}\]
\end{definition}

The set of all guarded terms with top forms a TopKAT:

\begin{definition}
    Given an alphabet \(K, B\),
    the \keyword{language TopKAT} or \(\topLangModel\) is a TopKAT where 
    \begin{itemize}[noitemsep]
        \item The elements are subsets of \(\topGuardedTerm\);
        \item The tests are subsets of \(\unitTopLangModel\);
        \item The multiplication is coalesced product;
        \item The addition is set union;
        \item The star operator is defined as follows:
        \[\starOf{S} = \bigunion_{n \in \nat} S^{n}
        \twhere S^{0} = \unitTopLangModel \tand S^{k + 1} = S \diamond S^{k};\] 
        \item The complement of test \(b\) is \(\unitTopLangModel - b\);
        \item the top element is the set \(\topGuardedTerm\);
        \item the additive identity is the empty set;
        \item the multiplicative identity is \(\unitTopLangModel\).
    \end{itemize}
\end{definition}

It is straightforward to show \(\topLangModel\) is a TopKAT.
Because it is the language KAT with alphabet \((K \union \{\top\}, B)\), 
thus it satisfy all the axioms of KAT. 
Because a TopKAT is just a KAT with a largest element,
and the largest element in \(\topLangModel\) 
is the set of all guarded terms: \(\topGuardedTerm\)
(with the order in this model, which is just subset ordering),
thus \(\topLangModel\) is a KAT.

\begin{definition}
    Given an alphabet \(K, B\), the \keyword{standard valuation} is
    \(g: K \union B \to \topLangModel\) defined as follows:
    \begin{align*}
        g(\synt{p}) & = \{\alpha \synt{p} \beta \mid \alpha \beta \in 1_{\languageModel}\} \\
        g(\synt{b}) & = \{\alpha \mid \text{\(\synt{b}\) appears positively in \(\alpha\)}\}
    \end{align*}
    We call the interpretation \(\interpWith{g}{-}: \TKATTerm{K}{B} \to \topLangModel\)
    the \keyword{standard interpretation}.
\end{definition}

Let's also recall some definitions regarding the standard interpretation of KAT:
\begin{definition}[\cite{goos_kleene_1997}]
  The function \(G\) takes a KAT term with alphabet \((K, B)\) and returns
  its standard interpretation as a guarded term in KAT.  
    And \(G\) is defined inductively as follows:
    \begin{align*}
        G(0) & \definedAs \emptyset \\
        G(1) & \definedAs \unitTopLangModel\\
        G(\synt{p}) & \definedAs \interpWith{g}{\synt{p}} 
            & \synt{p} \in K\\
        G(\synt{b}) & \definedAs \interpWith{g}{\synt{b}} 
            & \synt{b} \in B\\
        G(\synt{t_1} + \synt{t_2}) & \definedAs G(\synt{t_1}) \union G(\synt{t_2}) \\
        G(\synt{t_1} \synt{t_2}) & \definedAs 
            G(\synt{t_1}) \diamond G(\synt{t_2}) \\
        G(\starOf{\synt{t}}) & \definedAs
            \bigunion_{n \in \nat} {(G(\synt{t}))}^{n} 
    \end{align*}
\end{definition}

It is important for later development to note that the standard 
interpretation of a KAT term is constructed in the exact same way 
as a TopKAT term, 
the only difference is that it is missing the \(\top\) case.
Given an alphabet \(K, B\),
we notice that all the terms in \(\TKATTerm{K}{B}\) can be seen as KAT terms over the alphabet 
\((K \union \{\top\}, B)\).
Thus the function \(G\) can be applied onto elements of \(\TKATTerm{K}{B}\),
where it will simply regard \(\top\) as a action primitive, 
instead of the top element. 

The strategy to prove completeness and decidability of TopKAT with language model 
is by reducing all TopKAT terms into KAT terms.
For each TopKAT term, we can construct a KAT term that is equivalent to it.  
Since KAT equivalence is subsumed by TopKAT equivalence and KAT equivalence 
is known to be decidable in PSPACE, 
we have completeness and decidability of TopKAT equivalence.

\begin{theoremEnd}{lemma}\label{the: topkat complete core lemma}
Given an alphabet \((K, B)\),
there exists a function \(r: \TKATTerm{K}{B} \to \TKATTerm{K}{B}\),
s.t.\ for all terms \(\synt{t} \in \TKATTerm{K}{B}\)
\begin{itemize}[noitemsep]
    \item \(\validUnderModels{\allTopKATs}{\synt{t} = r(\synt{t})}\)
    \item \(G(r(\synt{t})) = \interpWith{g}{\synt{t}}\)
\end{itemize}
where \(r(\synt{t})\) simply replaces all the \(\top\) in \(\synt{t}\)
with \(\starOf{(\sum K + \top)}\).
\end{theoremEnd}

\begin{proofEnd}
    This can be shown with structural induction on \(\synt{t}\).
    As the definition of \(\interpWith{g}{-}\) and \(G\) coincides on all cases except 
    the \(\top\) case and we only change the \(\top\) symbols in \(\synt{t}\),
    all the other cases are trivial except the \(\top\) case.

    \itemTitle{the \(\top\) case:}
    we first show \[\interpWith{g}{\top} = G\left(\starOf{(\sum K + \top)}\right),\]
    which by definition of interpretation of \(\top\) is just 
    \[\{\alpha_{0} \synt{p}_1 \alpha_1 \dots \synt{p}_n \alpha_n \mid
    \alpha_{i} \in \unitTopLangModel, \synt{p}_i \in K \union \{\top\}, n \in \nat\} 
    = G\left(\starOf{(\sum K + \top)}\right)\]
    We observe that the following fact can be proven by induction on \(n\): 
    \[G\left((\sum K + \top)^{n}\right) = 
    \{\alpha_{0} \synt{p}_1 \alpha_1 \dots \synt{p}_n \alpha_n \mid
    \alpha_{i} \in \unitTopLangModel, \synt{p}_i \in K \union \{\top\}\},\]
    By definition of language KAT \cite[Section~3]{goos_kleene_1997} 
    \begin{align*}
        G\left(\starOf{(\sum K + \top)}\right) 
        & = \bigunion_{n \in \nat} G\left((\sum K + \top)^{n}\right) \\
        & = \bigunion_{n \in \nat} 
            \{\alpha_{0} \synt{p}_1 \alpha_1 \dots \synt{p}_n \alpha_n \mid
            \alpha_{i} \in \unitTopLangModel, \synt{p}_i \in K \union \{\top\}\} \\
        & = \{\alpha_{0} \synt{p}_1 \alpha_1 \dots \synt{p}_n \alpha_n \mid
            \alpha_{i} \in \unitTopLangModel, \synt{p}_i \in K \union \{\top\}, n \in \nat\} \\
        & = \interpWith{g}{\top}
    \end{align*}

    Then we show \(\validUnderModels{\allTopKATs}{\top = \starOf{(\sum K + \top)}}.\)
    \begin{itemize}
        \item we show 
            \(\validUnderModels{\allTopKATs}{\top \geq \starOf{(\sum K + \top)}}\),    
            by the axiom that \(\top\) is the largest element 
        \item we show 
            \(\validUnderModels{\allTopKATs}{\top \leq \starOf{(\sum K + \top)}}\), 
            by unfolding \(\starOf{(\sum K + \top)}\) twice:
            \[\validUnderVal{\allTopKATs}{
                \starOf{(\sum K + \top)} 
                = 1 + (\sum K + \top) + (\sum K + \top)^2 \starOf{(\sum K + \top)}
                \geq \top
            }.\]
    \end{itemize}

    We also show couple other cases to give the reader a intuition on why 
    they are trivial.

    \itemTitle{Primitive Action (not \(\top\)) Case:}
    For any primitive action \(\synt{p}\), because \(r(\synt{p}) = \synt{p}\),
    \[\validUnderModels{\allTopKATs}{\synt{p} = r(\synt{p})};\]
    and because \(G(\synt{p}) = \interpWith{g}{\synt{p}}\),
    \[G(r(\synt{p})) = G(\synt{p}) = \interpWith{g}{\synt{p}}.\]

    \itemTitle{Multiplication Case:}
    We first notice that by definition of \(r\), 
    \[r(\synt{t_1}) \cdot r(\synt{t_2}) = r(\synt{t_1} \cdot \synt{t_2}).\]

    Given the induction hypothesis 
    \(\validUnderModels{\allTopKATs}{\synt{t_1} = r(\synt{t_1})}\)
    and \(\validUnderModels{\allTopKATs}{\synt{t_2} = r(\synt{t_2})}\),
    \[\validUnderModels{\allTopKATs}
    {\synt{t_1} \cdot \synt{t_2} = r(\synt{t_1}) \cdot r(\synt{t_2}) 
    = r(\synt{t_1} \cdot\synt{t_2})}.\]
    We also have the induction hypothesis \(G(r(\synt{t_1})) = \interpWith{\synt{t_1}}{g}\)
    and \(G(r(\synt{t_2})) = \interpWith{\synt{t_2}}{g}\),
    then 
    \[G(r(\synt{t_1} ~ \synt{t_2}))) 
    = G(r(\synt{t_1})) \cdot G(r(\synt{t_2})) 
    = \interpWith{g}{\synt{t_1}} \cdot \interpWith{g}{\synt{t_2}}
    = \interpWith{g}{\synt{t_1 ~ t_2}}\]
\end{proofEnd}

\begin{theoremEnd}[normal]{theorem}[Completeness of the standard interpretation]\label{the: topkat complete}
    Given an alphabet \(K, B\), and two TopKAT terms \(\synt{t_1}, \synt{t_2} \in \TKATTerm{K}{B}\) 
    The following conditions are equivalent
    \begin{itemize}
        \item \(\validUnderModels{\allTopKATs}{\synt{t_1} = \synt{t_2}}\)
        \item \(\validUnderVal{g}{\synt{t_1 = t_2}}\)
        \item \(\validUnderModels{\allKATs}{r(\synt{t_1}) = r(\synt{t_1})}\)
    \end{itemize}
\end{theoremEnd}

\begin{proofEnd}
\begin{itemize}
    \item 
    First we show that 
    \(\validUnderVal{g}{\synt{t_1 = t_2}} \iff \validUnderModels{\allKATs}{r(\synt{t_1}) = r(\synt{t_2})}:\)

    \(\impliedby\), because all TopKATs are KATs, we have 
    \[\validUnderModels{\allTopKATs}{r(\synt{t_1}) = r(\synt{t_2})}\]
    by \Cref{the: topkat complete core lemma},
    \[
        \validUnderModels{\allTopKATs}{r(\synt{t_1}) = \synt{t_1}} \tand
        \validUnderModels{\allTopKATs}{r(\synt{t_2}) = \synt{t_2}}.
    \]
    we have 
    \[\validUnderModels{\allTopKATs}{\synt{t_1} = \synt{t_2}}.\]
    and since \(g\) is a TopKAT valuation, 
    we have \[\validUnderVal{g}{\synt{t_1 = t_2}}.\]
    then \(\implies\), 
    by \Cref{the: topkat complete core lemma}
    \[
        G(r(\synt{t_1})) = \interpWith{g}{\synt{t_1}} \tand
        G(r(\synt{t_2})) = \interpWith{g}{\synt{t_2}}
    \]
    because \(\validUnderVal{g}{\synt{t_1} = \synt{t_2}}\),
    then \(\interpWith{g}{\synt{t_1}} = \interpWith{g}{\synt{t_2}}\),
    thus \[G(r(\synt{t_1})) = \interpWith{g}{\synt{t_1}} = 
    \interpWith{g}{\synt{t_2}} = G(r(\synt{t_2})).\]
    Therefore by completeness over the standard interpretation of 
    KAT~\cite[Theorem~8]{goos_kleene_1997},
    we have \(\validUnderModels{\allKATs}{r(\synt{t_1}) = r(\synt{t_2}})\),
    
    \item Then we show 
    \[\validUnderVal{g}{\synt{t_1 = t_2}} \iff \validUnderModels{\allTopKATs}{\synt{t_1} = \synt{t_2}}\]
    \(\impliedby\) is trivial, since \(g\) is a TopKAT interpretation.
    \(\implies\) is shown as follows:
    we have \[\validUnderVal{g}{\synt{t_1 = t_2}} \iff \validUnderModels{\allKATs}{r(\synt{t_1}) = r(\synt{t_1})}\]
    since every TopKAT is a KAT, 
    therefore \[\validUnderModels{\allTopKATs}{r(\synt{t_1}) = r(\synt{t_2})}.\]
    Since \(\validUnderModels{\allTopKATs}{r(\synt{t_1}) = \synt{t_1}}\),
    \(\validUnderModels{\allTopKATs}{r(\synt{t_2}) = \synt{t_2}}\),
    we have \(\validUnderModels{\allTopKATs}{\synt{t_1} = \synt{t_2}}\).
\end{itemize}

\end{proofEnd}

\begin{theoremEnd}[normal]{corollary}\label{the: PSPACE for TopKAT equalities}
    Deciding equalities of an arbitrary equality in TopKAT is PSPACE-complete.
\end{theoremEnd}

\begin{proofEnd}
    By \Cref{the: topkat complete}, we have 
    \[\validUnderModels{\allKATs}{r(\synt{t_1}) = r(\synt{t_1})} 
    \iff \validUnderModels{\allTopKATs}{\synt{t_1} = \synt{t_2}}.\]
    By construction of \(r\), 
    the size of \(r(\synt{t})\) is polynomial in the 
    size of \(\synt{t}\) plus the size of the alphabet.
    Then if we constrain the alphabet to only include primitives that
    appeared in \(\synt{t_1}\) or \(\synt{t_2}\),
    then the size of \(r(\synt{t_1})\) and \(r(\synt{t_2})\) 
    will be polynomial in the size of \(\synt{t_1}\) plus \(\synt{t_2}\).

    Since KAT equality is decidable in PSPACE,
    and \(r(\synt{t_1}), r(\synt{t_2})\) only take polynomial space to store,
    we can first compute and store \(r(\synt{t_1}), r(\synt{t_2})\),
    then decide their equality as two KAT terms. 
    This algorithm will be in PSPACE.

    Consider two terms \(\synt{t_1}, \synt{t_2} 
    \in \KATTerm{K}{B} \subseteq \TKATTerm{K}{B}\), 
    \[\validUnderModels{\allTopKATs}{\synt{t_1} = \synt{t_2}} 
    \iff \validUnderModels{\allKATs}{\synt{t_1} = \synt{t_2}}.\]
    Thus deciding TopKAT equalities should be at least as hard as KAT equalities. 
    Because deciding KAT equalities are known to be 
    PSPACE-complete~\cite{cohen_complexity_1999},
    thus deciding TopKAT quality is PSPACE-hard.

    Thus, deciding TopKAT equalities is PSPACE-hard and can be decided in PSPACE,
    deciding TopKAT equalities is PSPACE-complete.
\end{proofEnd}

Besides language TopKATs and relational TopKATs, 
we have already mentioned a more general class of relation-based TopKATs in the proof of
\Cref{the: topkatstar incomplete over relational model},
where the top element is not necessarily the complete relation.
The motivation to investigate this class of TopKATs is more than simple mathematical curiosity:
being a relation-based class of TopKATs, 
these structures have the potential to model programs as a input/output relations, 
just like relational TopKATs.

\begin{definition}
    A \keyword{general relational TopKAT} is a relational TopKAT where 
    the top element is not necessarily the complete relation.
    We denote all the general relational TopKATs \(\allGRELs\).
\end{definition}
Because composition distributes over infinite unions of relations, we can show the
\(\star\)-continuity axiom holds in \(\allGRELs\).  Thus, all general relational
TopKATs are TopKATs.

\begin{example}\label{exp: GREL domain example}
The following elements form a general relational TopKAT over \(\{0, 1\}\),
but not a relational TopKAT\@:
\begin{align*}
    & \emptyset \\ 
    & \{(1, 1)\} \\
    & \{(0, 1)\} \\
    & \{(0, 0), (1, 1)\} \\
    & \{(0, 1), (0, 0), (1, 1)\}
\end{align*}
where the top element is \(\{(0, 1), (0, 0), (1, 1)\}\),
not the complete relation on \(\{0, 1\}\)
\end{example}

We can extend some definitions about relational TopKATs to general relational TopKATs.
\begin{definition}
Given an alphabet \(K, B\)
\begin{itemize}
    \item for all general relational TopKATs \(\algebra{R}\), 
    a \keyword{general relational valuation} is a function 
    \[u: K \union B \to \algebra{R};\]
    \item A \keyword{general relational interpretation} 
    \[\interpWith{u}{-}: \TKATTerm{K}{B} \to \algebra{R}\]
    is the interpretation generated by \(r\) as in \Cref{def:kat-valuation};
    \item For two terms \(t_1, t_2 \in \TKATTerm{K}{B}\),
    the statement \(t_1 = t_2\) is \keyword{valid under all general relational interpretations}
    if for all general relational valuations \(u\):
    \[\interpWith{u}{t_1} = \interpWith{u}{t_2},\]
    we write it as \[\validUnderModels{\allGRELs}{t_1 = t_2}.\]
\end{itemize}
\end{definition}

\begin{theoremEnd}[normal]{theorem}[Completeness of \(\allGRELs\)]\label{the: topkatstar complete over GREL}
Given an alphabet \(K, B\), and two TopKAT terms \(\synt{t_1, t_2} \in \TKATTerm{K}{B}\),
\[\validUnderModels{\allGRELs}{\synt{t_1 = t_2}} \iff \validUnderModels{\allTopKATs}{\synt{t_1 = t_2}}\]
\end{theoremEnd}

\begin{proofEnd}
Same proof as in~\cite[Lemma 5, Theorem 6]{goos_kleene_1997},
we define the following injective homomorphism from a language TopKAT 
to a general relational TopKAT
\[h(S) = \{(s_1, s_1 \diamond s) \mid s_1 \in \topGuardedTerm, s \in S\}\]

We first verify that it is homomorphism, 
most of the cases is the same as~\cite{goos_kleene_1997},
the only non-trivial case is to show that \(h(\topGuardedTerm)\) is the largest element,
which can be proven just by unfolding the definitions. 

Proving \(\impliedby\):
Because every general relational TopKAT is a TopKAT.
If a statement is true for all TopKAT, it is true for all general relational TopKATs.

Proving \(\implies\): 
since \(h\) is an injective homomorphism, the domain is isomorphic to its range.
Thus for all language TopKATs, there exists an isomorphic general relational TopKAT\@.
If \(\synt{t_1} = \synt{t_2}\) in all general relational TopKATs,
then \(\synt{t_1} = \synt{t_2}\) is also true in all language TopKATs.
Finally by completeness over the standard interpretation,
we have \(\allTopKATs\) is complete over \(\allGRELs\).
\end{proofEnd}

\subsection{Equational Expressiveness of General Relational TopKATs}

Given the completeness of general relational TopKAT, it is natural to wonder
whether we can encode incorrectness logic in \(\allGRELs\), so that the
incompleteness of \(\allTopRELs\) is no longer a problem to reason about
incorrectness logic in the theory of TopKAT\@.

However, we notice the formulation of codomain will no longer work in \(\allGRELs\).
Recall the formulation of domain in \(\allTopRELs\):
\[\top p = \top q \iff \codomain{p} = \codomain{q}.\]
We take the \(\allTopKATs\) in \Cref{exp: GREL domain example},
and let \(p = \{(0, 1)\}, q = \{(1, 1)\}\), then we have \(\codomain{p} = \codomain{q}\)
but 
\begin{align*}
\top p &= \{(0, 1), (0, 0), (1, 1)\} \thenCompo \{(0, 1)\} = \{(0, 1)\}\\
\top q &= \{(0, 1), (0, 0), (1, 1)\} \thenCompo \{(1, 1)\} = \{(0, 1), (1, 1)\},
\end{align*}
hence \(\top p \neq \top q\), 
thus \(\codomain{p} = \codomain{q} \implies \top p = \top q\) no longer holds.

Using the same method, 
we can also show that the formulation of incorrectness triple no longer holds in general relational TopKATs.
Let \(b = \{(0, 0), (1, 1)\}, p = \{(0, 1)\}, c = \{(1, 1)\}\),
then 
\begin{itemize}
\item the incorrectness triple \(\incorTriple{b}{p}{c}\) holds.
\item \(\top b p = \top p = \{(0, 1)\}\) but \(c = \{(1, 1)\}\), 
    hence \(\top b p \not \geq c\).
\end{itemize}
Therefore \(\top b p \geq c \iff \incorTriple{b}{p}{c}\) no longer holds in general relational TopKATs.

It would be interesting to see if there exists another way to express
incorrectness logic in general relational TopKATs.  Unfortunately, we will see
that not only it is not possible to express incorrectness in general relational
TopKATs, but general relational TopKATs have the same equational expressiveness
as \(\allRELs\): any predicate expressible in general relational TopKATs can already be
expressed in $\allRELs$.

\begin{definition}[Equational Expressiveness of TopKATs]
  Given an alphabet \(K, B\), a TopKAT \(\algebra{K}\), and an \(n\)-ary
  predicate \(P: \algebra{K}^{n} \to \bool\),
  we say two terms \(\synt{t_1}, \synt{t_2} \in \TKATTerm{K}{B}\)
  \keyword{express} the predicate \(P\) over primitives 
  \(\synt{p_1}, \dots, \synt{p_n} \in K \union B\) in \(\algebra{K}\),
  if for all valuations \(u: K \union B \to \algebra{K}\):
  \[\validUnderVal{u}{\synt{t_1} = \synt{t_2}} \iff 
  P(\interpWith{u}{\synt{p_1}}, \interpWith{u}{\synt{p_2}}, \dots \interpWith{u}{\synt{p_n}})\]

  A predicate is \keyword{expressible in \(\allGRELs\)} if there
  exists a pair of TopKAT terms that express the predicate in all
  general relational TopKATs.
\end{definition}

Then we show that \(\allGRELs\) has the same equational expressiveness as \(\allRELs\).
Intuitively, the proof of equiexpressiveness
exploits the fact that we can ``simulate'' the \(\top\) term using 
the star of the sum of the entire alphabet.
Thus, given two TopKAT terms that can express a predicate, 
we can construct two KAT terms, where \(\top\) is simulated as above,
to express the same predicate.

\begin{theoremEnd}{lemma}\label{the: grel formulatablity imply rel formulatablity lemma}
    Given an alphabet \(K, B\),
    and a term \(\synt{t} \in \TKATTerm{K}{B}\),
    there exists a term \(\synt{\hat{t}} \in \KATTerm{K}{B}\),
    s.t.\ for all relational KATs \(\algebra{R}\) over \(X\) and
    relational valuations \(u: K \union B \to \algebra{R}\),
    there exists a general relational TopKAT \(\algebra{\hat{R}}\) over \(X\)
    and valuation \(\hat{u}: K \union B \to \algebra{\hat{R}}\) that is point-wise equal to \(u\),
    i.e. \[\forall \synt{p} \in K \union B,~ \hat{u}(\synt{p}) = u(\synt{p}) \]
    s.t.\ the following hold
    \[\interpWith{\hat{u}}{\synt{t}} = \interpWith{u}{\hat{\synt{t}}}\]
\end{theoremEnd}

\begin{proofEnd}
    We first state our construction.
    \begin{itemize}
        \item For all terms \(\synt{t} \in \TKATTerm{K}{B}\), 
            we construct \(\synt{\hat{t}} \in \KATTerm{K}{B}\) via replacing all 
            the \(\top\) symbols with the following term: 
            \[\starOf{(\sum K)} \]
            this is a valid term, since \(K\) is finite. 
        \item For all relational KATs \(\algebra{R}\), 
            we construct the following general relational TopKATs \(\hat{\algebra{R}}\),
            s.t.\ it is generated by the following set with composition, union, and transitive closure,
            which corresponds to the multiplication, addition, and star operation in \(\allTopKATs\):
            \[\{u(\synt{p}) \mid \synt{p} \in K \union B\},\]
            This \(\allTopKATs\) is indeed a general relational TopKAT, because it has a top element,
            which is the reflexive transitive closure of all the elements
            (the proof of why this is the largest element will be shown later):
            \[
                \bigunion_{n \in \nat} {(\bigunion \{u(\synt{p}) \mid \synt{p} \in K\})}^{n}
            \]
            where for all sets of relations \(R\), 
            its natural number exponential \(R^{n}\) is inductively defined as follows:
            \begin{align*}
                R^{0} & \definedAs \{(x, x) \mid x \in X\}, \\
                R^{k + 1} & \definedAs \{r \thenCompo r' \mid r \in R, r' \in R^{k}\}.
            \end{align*}
    \end{itemize}
    
    For all terms \(\synt{t} \in \TKATTerm{K}{B}\), and all valuations \(u\)
    we can prove by induction that structure of \(\synt{t}\) that
    \[\interpWith{\hat{u}}{\synt{t}} = \interpWith{u}{\hat{\synt{t}}}.\]
    Because \(\synt{t}\) and \(\synt{\hat{t}}\) are the same except where \(\top\) occurs,
    and because \(u\) and \(\hat{u}\) are point-wise equal,
    then the only non-trivial case is the top case:
    where we need to show 
    \[\interpWith{\hat{u}}{\top} = \interpWith{u}{\starOf{(\sum K)} }.\]
    By definition of relational KATs, we have:
    \[
        \interpWith{u}{\starOf{(\sum K)} } = 
        \bigunion_{n \in \nat} {(\bigunion \{u(\synt{p}) \mid \synt{p} \in K\})}^{n}
    .\]
    Then we need to show this element is the largest element in \(\algebra{\hat{R}}\).
    For simplicity, we will denote this element as \({\top}_\algebra{\hat{R}}\).

    Since \(\algebra{\hat{R}}\) is generated by composition, union and reflexive closure 
    of the following
    \[\{u(\synt{p}) \mid \synt{p} \in K \union B\}\]
    then we can prove this by induction on the generation of elements in \(\algebra{\hat{R}}\).
    \begin{itemize}
        \item \itemTitle{Base case:} 
        \begin{itemize}
            \item for an element \(q \in \{u(\synt{p}) \mid \synt{p} \in K\}\)
                \[q \subseteq {\{\bigunion u(\synt{p}) \mid \synt{p} \in K\}}^{1} 
                \subseteq {\top}_\algebra{\hat{R}}\]
                for element \(a \in \{u(\synt{b}) \mid \synt{b} \in B\}\)
                \[a \subseteq{\{\bigunion u(\synt{p}) \mid \synt{p} \in K\}}^{0} 
                \subseteq {\top}_\algebra{\hat{R}}\]
        \end{itemize}
        \item \itemTitle{Composition case:}
            assume for two elements \(p, q \in \algebra{\hat{R}}\), 
            \(p, q \subseteq {\top}_\algebra{\hat{R}}\),
            we need to show \(p \thenCompo q \subseteq {\top}_\algebra{\hat{R}}\).
            This result is true because that \({\top}_\algebra{\hat{R}}\) is a transitive closure.
            Therefore \[{\top}_\algebra{\hat{R}} {\top}_\algebra{\hat{R}} = {\top}_\algebra{\hat{R}}\]
            By multiplication preserves order in \(\allTopKATs\)
            \[p \thenCompo q = p q \leq p {\top}_\algebra{\hat{R}} 
            \leq {\top}_\algebra{\hat{R}} {\top}_\algebra{\hat{R}} = {\top}_\algebra{\hat{R}}\]
        \item \itemTitle{Union case:}
            assume for two elements \(p, q \in \algebra{\hat{R}}\).
            by induction hypothesis, \(p \subseteq \top_{\algebra{\hat{R}}}\)
            and \(q \subseteq \top_{\algebra{\hat{R}}}\),
            by property of set we have \(p \union q \subseteq {\top}_\algebra{\hat{R}}\).
        \item \itemTitle{Star case:}
            assume \(p \subseteq {\top}_\algebra{\hat{R}}\)
            Then we need to show \(\starOf{p} \subseteq {\top}_\algebra{\hat{R}}\).
            Since \({\top}_\algebra{\hat{R}}\) is a transitive closure, 
            we can show the following:
            \[\starOf{{\top}_\algebra{\hat{R}}} = {\top}_\algebra{\hat{R}}\]
            Thus, because \(p \subseteq {\top}_\algebra{\hat{R}}\) and star preserve order 
            in \(\allTopKATs\), we have 
            \[\starOf{p} \subseteq \starOf{{\top}_\algebra{\hat{R}}} = {\top}_\algebra{\hat{R}}\]
        \item \itemTitle{Complement case:}
            Since complement only defined on the boolean sub-algebra, 
            and results in an element of the boolean sub-algebra.
            Therefore for all \(b\) in the boolean sub-algebra
            \[\compl{b} \subseteq 1 = {\{\bigunion u(\synt{p}) \mid \synt{p} \in K\}}^{0} 
            \subseteq {\top}_\algebra{\hat{R}},\]
            where the 1 is the identity relation.
    \end{itemize}
\end{proofEnd}

\begin{theoremEnd}[normal]{theorem}[Equational Expressiveness of General Relational TopKATs]\label{the: GREL expresses same predicate as REL KAT}
    Given an alphabet \(K, B\), an \(n\)-ary predicate \(P\),
    the predicate \(P\) over primitives \(\synt{p_1}, \synt{p_2}, \dots, \synt{p_n}\)
    is expressible in \(\allGRELs\) iff it is expressible in \(\allRELs\).
\end{theoremEnd}

\begin{proofEnd}
  Since general relational TopKATs are a subclass of \(\allRELs\), if two KAT
  terms \(\synt{t_1} = \synt{t_2}\) express a predicate in \(\allRELs\), the
  same pair of terms will express the same predicate in general relational
  TopKATs.

  Then we show the other direction: if the predicate is expressible in general
  relational TopKATs, then it is expressible in \(\allRELs\).  By \Cref{the:
    grel formulatablity imply rel formulatablity lemma}, if
  \(\synt{t_1} = \synt{t_2}\) express the predicate in \(\allGRELs\), we
  construct \(\hat{\synt{t_1}}\) and \(\hat{\synt{t_2}}\); and for all
  relational valuations \(u: K \union B \to \algebra{R}\), we find the general
  relational valuation \(\hat{u}: K \union B \to \algebra{\hat{R}}\).  Since
  \(\synt{t_1} = \synt{t_2}\) expresses \(P\) over
  \(\synt{p_1}, \synt{p_2}, \dots, \synt{p_n}\) in \(\allGRELs\), we have
    \[\validUnderVal{\hat{u}}{\synt{t_1} = \synt{t_2}} \iff 
    P(\interpWith{\hat{u}}{\synt{p_1}}, \interpWith{\hat{u}}{\synt{p_2}}, \dots \interpWith{\hat{u}}{\synt{p_n}}).\]
    By \Cref{the: grel formulatablity imply rel formulatablity lemma},
    we know that 
    \[\forall \synt{p} \in K \union B, ~ \interpWith{\hat{u}}{\synt{p}} = \interpWith{u}{\synt{p}},\]
    thus 
    \[
        P(\interpWith{\hat{u}}{\synt{p_1}}, \interpWith{\hat{u}}{\synt{p_2}}, \dots \interpWith{\hat{u}}{\synt{p_n}})
        \iff 
        P(\interpWith{u}{\synt{p_1}}, \interpWith{u}{\synt{p_2}}, \dots \interpWith{u}{\synt{p_n}}).
    \]
    Also by \Cref{the: grel formulatablity imply rel formulatablity lemma},
    we have \(\interpWith{\hat{u}}{\synt{t_1}} = \interpWith{u}{\synt{\hat{t_1}}}\)
    and \(\interpWith{\hat{u}}{\synt{t_2}} = \interpWith{u}{\synt{\hat{t_2}}}\),
    hence
    \[\validUnderVal{u}{\synt{\hat{t_1} = \hat{t_2}}} \iff 
    \validUnderVal{\hat{u}}{\synt{t_1 = t_2}}.\]
    Finally, we conclude: for all relational valuations \(u\),
    \[
        \validUnderVal{u}{\synt{\hat{t_1} = \hat{t_2}}} 
        \iff 
        \validUnderVal{\hat{u}}{\synt{t_1 = t_2}}
        \iff 
        P(\interpWith{\hat{u}}{\synt{p_1}}, \interpWith{\hat{u}}{\synt{p_2}}, \dots \interpWith{\hat{u}}{\synt{p_n}})
        \iff 
        P(\interpWith{u}{\synt{p_1}}, \interpWith{u}{\synt{p_2}}, \dots \interpWith{u}{\synt{p_n}})
    \]
    Therefore \(\synt{\hat{t_1} = \hat{t_2}}\) expresses the predicate 
    \(P\) over \(\synt{p_1}, \synt{p_2}, \dots, \synt{p_n}\) in \(\allRELs\).
\end{proofEnd}

\begin{theoremEnd}[normal]{corollary}\label{the: GREL cannot formulate incorrectness logic}
    General relational TopKATs cannot express incorrectness logic.
\end{theoremEnd}

\begin{proofEnd}
    By \Cref{the: KAT not able to express incorrectness logic} 
    we know that \(\allRELs\) cannot express incorrectness logic,
    then by \Cref{the: GREL expresses same predicate as REL KAT},
    we know that all the predicates that cannot be expressed in \(\allRELs\)
    cannot be expressed in general relational TopKATs.

    Therefore general relational TopKATs cannot express incorrectness logic.
\end{proofEnd}


\section{Modeling Errors in Incorrectness Logic}\label{sec: failkat and error}

One of the advantages of algebraic methods is the ease of extension.
In this section, we show how extending TopKAT with failure can naturally give 
rise to incorrectness triples that express abnormal termination.  
The main difference arises from the short-circuiting behavior of sequencing with errors.
This can be seen in the following rule for sequential composition, which states
that if an error already occurred in \(p\), \(q\) will not be executed.
\[
\inferrule[Composition-Fail]
{ \incorTriple{b}{p}{\errState: c}}
{ \incorTriple{b}{pq}{\errState: c}}.
\]

To capture this type of control flow, we adapt the ideas from
\citet[Definition~3]{esparza_equational_2017}, who investigated similar issues
in the setting of KAT\@.
\begin{definition}[FailTopKAT]
  A FailTopKAT is a tuple \((\algebra{F},\algebra{K},\algebra{B},\fail)\), where
  \((\algebra{K}, \algebra{B})\) is a TopKAT, \(\fail \in \algebra{F}\), and
  \(\algebra{K} \subsetneq \algebra{F}\). The set \(\algebra{F}\) has the
  structure of a KAT that extends that of \(\algebra{K}\), except that the right
  annihilation rule \(p \cdot 0 = 0\) need not hold.  Instead,
  \[\fail \cdot p = \fail,\] where \(p\) is any element of \(\algebra{F}\).
  (Crucially, we do \emph{not} assume \(\top \geq \fail\).) We call
  \((\algebra{K},\algebra{B})\) the \keyword{fail-free subalgebras}, 
  which model programs that do not fail.  
  We will omit some of \(\algebra{K},\algebra{B},\fail\),
  if they are not used or can be inferred from the context.

  The class of all FailTopKATs is denoted \(\allFailTopKATs\).
\end{definition}
Note that the original definition of \citet{esparza_equational_2017} allows for
try-catch statements and different types of errors.  We omit these features for
simplicity, since they are not needed in incorrectness
logic~\cite{ohearn_incorrectness_2020}.  \iffull For a more explicit definition
of \(\allFailTopKATs\) with all the rules, please refer to the
\hyperref[def: explicit fail top kat]{definition} on page~\pageref{def: explicit
  fail top kat}.  \fi

There exists a canonical procedure for extending a TopKAT \(\algebra{K}\) with
failures.  The idea, which we adapt from Construction
F~\cite[Definition~4]{esparza_equational_2017} is to consider elements of the
form \((p,p') \in \algebra{K} \times \algebra{K}\), where \(p\) represents
executions that terminate normally, and \(p'\) represents executions that fail.
\begin{definition}[Construction F for \(\allFailTopKATs\)]
  Given a TopKAT \((\algebra{K}, \algebra{B})\), we construct a FailTopKAT
  \((\algebra{F}, \algebra{K'}, \algebra{B'})\).  The carrier sets are defined
  as
  \begin{align*}
    \algebra{F} & \definedAs \algebra{K} \times \algebra{K} &
    \algebra{K'} & \definedAs \algebra{K} \times \{0_{\algebra{K}}\} &
    \algebra{B'} & \definedAs \algebra{B} \times \{0_{\algebra{K}}\},
  \end{align*}
  where \(0_{\algebra{K}}\) is the additive identity in \(\algebra{K}\).  The
  operations of \(\algebra{F}\) are defined as follows
  \begin{align*}
      0_{\algebra{F}} & \definedAs (0_{\algebra{K}}, 0_{\algebra{K}}) \\
      1_{\algebra{F}} & \definedAs (1_{\algebra{K}}, 0_{\algebra{K}}) \\
      \top_{\algebra{F}} & \definedAs (\top_{\algebra{K}}, 0_{\algebra{K}}) \\
    (p, p') (q, q') & = (p q, p' + pq') \\
    (p, p') + (q, q') & = (p + q, p' + q') \\
    \starOf{(p, p')} & = (\starOf{p}, (\starOf{p}) p') \\
    \fail & = (0_{\algebra{K}}, 1_{\algebra{K}}) \\
    \compl{(b, 0_{\algebra{K}})} & = (\compl{b}, 0_{\algebra{K}}).
  \end{align*}
\end{definition}
To develop some intuition for this construction, suppose that \(\algebra{K}\) is a
relational TopKAT over \(X\).  There exists a canonical embedding
of \(\algebra{F}\) in \(\powerSet{X \times X \times \{\okState,\errState\}}\) that identifies
\((p,p') \in \algebra{F}\) with the relation
\(r = p \times \{\okState\} \cup p' \times \{\errState\} \subseteq X \times X \times \{\okState,\errState\}\).
Intuitively, \((x,y,\epsilon) \in r\) means that a program took the input state \(x\) to
the output state \(y\), and the bit \(\epsilon \in \{\okState,\errState\}\) signals whether an error has
occurred.  By looking at the definition of sequential composition under this
reading, it says that we get an error either by getting an error when running
the first command (\(p'\)), or if we successfully run the first command, but get
an error when running the second (\(pq'\));
and we terminates normally only we sequentially execute \(p\) and then \(q\).

The semantics of~\citet[Fig.~4]{ohearn_incorrectness_2020} follows the same
pattern, except that he considered the cases \(\okState\) and \(\errState\) in
separate relations.  We can merge them back into a tuple, for example the
semantics of \(\comSkip\)
\[(\{(x, x) \mid x \in X\}, \emptyset)\] 
coincides with the multiplicative identity \(1\) of 
applying construction F to a relational TopKAT \(\algebra{K}\): 
\[(1_{\algebra{K}}, 0).\]
In the same way, the sequential composition is multiplication,
the choice operator is addition, the star operator is the Kleene star,
and the \(\comError\) command is \(\fail\).
Thus applying F construction on a relational TopKAT will
capture the semantics of programs with abnormal termination:

\begin{definition}[Relational FailTopKAT]\label{def:relational-failtopkat}
  A relational FailTopKAT is a FailTopKAT constructed by applying construction F
  to a relational TopKAT\@. The class of all relational FailTopKATs is denoted
  \(\allFailTopRELs\).
\end{definition}

To better understand how to encode an incorrectness triple using FailTopKAT, we
propose a definition of incorrectness triple equivalent to the original one
\cite[Definition 1 and 4]{ohearn_incorrectness_2020}:
\begin{definition}\label{def: relational validity of incor triple with fail}
  Given a relational FailTopKAT \((\algebra{F}, \algebra{B})\),
  \(p \in \algebra{F}\), and \(b, c \in \algebra{B}\), for an error code
  \(\epsilon \in \{\okState, \errState\}\) an incorrectness triple
  \(\incorTriple{b}{p}{\epsilon: c}\) is valid if
    \[\codomain{b p} \supseteq \codomain{c \cdot \hat{\epsilon}}\] 
    where
    \[ \hat{\epsilon} \definedAs
      \begin{cases}
        1 & \text{if \(\epsilon = \okState\)} \\
        \fail & \text{if \(\epsilon = \errState\).}
      \end{cases}\]
    and the function \(\codomain{-}\) is extended entry-wise:
    \begin{align*}
        \codomain{(r, q)} & \definedAs (\codomain{r}, \codomain{q}).
    \end{align*}
\end{definition}

Following the development of \Cref{sec: formulating incorrectness logic}, we can
obtain a formulation of incorrectness triple with abnormal termination.

\begin{theoremEnd}{theorem}[Relational Validity]
  For all relational FailTopKATs \((\algebra{F}, \algebra{B})\), and \(p, q \in \algebra{F}\),
  we have
\begin{align*}
    \top p = \top q & \iff \codomain{p} = \codomain{q} \\
    \top p \leq \top q & \iff \codomain{p} \subseteq \codomain{q},
\end{align*}
\end{theoremEnd}

\begin{proofEnd}
  Write \(p = (p_1,p_2)\) and \(q = (q_1,q_2)\).  Then, by definition,
  \(\top p = (\top p_1, \top p_2)\) and \(\top q = (\top q_1, \top q_2)\).  Thus,
  \(\top p \leq \top q\) is equivalent to \(\top p_1 \leq \top q_1\) and
  \(\top p_2 \leq \top q_2\), which allows us to conclude.
\end{proofEnd}

\begin{theoremEnd}{corollary}\label{the: FailTopKAT can express incorrectness logic}
  For all relational FailTopKATs \(\algebra{F}\), for all \(p \in \algebra{F}\)
  and for all tests \(b, c \in \algebra{B}\), and for all
  \(\epsilon \in \{\okState,\errState\}\), the following holds
    \[\incorTriple{b}{p}{\epsilon: c} \iff \top b p \geq (c \cdot \hat{\epsilon}).\]
  where \(\hat{\epsilon}\) is defined in \Cref{def: relational validity of incor triple with fail}
\end{theoremEnd}

Thus, we can generalize incorrectness logic with errors to an arbitrary
FailTopKAT\@.

\begin{definition}[Abstract Incorrectness Triple With Failure]
  Given a \(\allFailTopKATs\) \((\algebra{F}, \algebra{K}, \algebra{B})\),
  \(b, c \in \algebra{B}\), \(p \in \algebra{F}\) and
  \(\epsilon \in \{\okState, \errState\}\), we define
  \[\incorTriple{b}{p}{\epsilon: c} \definedAs \top b p \geq c \cdot
    \hat{\epsilon}, \] where \(\hat{\epsilon}\) is defined as in 
    \Cref{def: relational validity of incor triple with fail}.
\end{definition}

\begin{figure}
    \begin{mathpar}
        \inferrule[Empty]
        {\\}{\incorTriple{b}{p}{\epsilon: 0}}

        \and
        \inferrule[Consequence]
        {b \leq b' \\ \incorTriple{b}{p}{\epsilon: c} \\ c' \leq c}
        {\incorTriple{b'}{p}{\epsilon: c'}}

        \and
        \inferrule[Disjunction]
        {\incorTriple{b_1}{p}{\epsilon: c_1} \\ \incorTriple{b_2}{p}{\epsilon: c_2}}
        {\incorTriple{b_1 + b_2}{p}{\epsilon: (c_1 + c_2)}}

        \and
        \inferrule[Identity]
        {\\} {\incorTriple{b}{1}{\okState: b, \errState: 0}}

        \and
        \inferrule[Composition-Fail]
        {\incorTriple{a}{p}{\errState: b}}
        {\incorTriple{a}{p q}{\errState: b}}

        \and
        \inferrule[Composition-Normal]
        {\incorTriple{a}{p}{\okState: b} \\ \incorTriple{b}{p}{\epsilon: c}}
        {\incorTriple{a}{pq}{\epsilon: c}}

        \and
        \inferrule[Choice-Left]
        {\incorTriple{b}{p}{\epsilon: c}}
        {\incorTriple{b}{p + q}{\epsilon: c}}

        \and
        \inferrule[Choice-Right]
        {\incorTriple{b}{q}{\epsilon: c}}
        {\incorTriple{b}{p + q}{\epsilon: c}}

        \and
        \inferrule[Assume]
        {\\}
        {\incorTriple{a}{b}{\okState: ab, \errState: 0}}

        \and
        \inferrule[Error]
        {\\}{\incorTriple{b}{\fail}{\errState: b}}

        \and
        \inferrule[Iter-Zero]
        {\\}{\incorTriple{b}{\starOf{p}}{\okState: b}}

        \and
        \inferrule[Iter-NonZero]
        {\incorTriple{b}{\starOf{p}p}{\epsilon: c}}
        {\incorTriple{b}{\starOf{p}}{\epsilon: c}}

        \and
        \inferrule[Iter-Dependent]
        {\forall n \in \nat,~ \incorTriple{b_{n}}{p}{\okState: b_{n+1}}}
        {\incorTriple{b_{0}}{\starOf{p}}{\okState: \sup_{n \in \nat}b_n}}
    \end{mathpar}
    \caption{Complete set of incorrectness logic proof rule with both normal and
      abnormal termination}\label{fig: all proof rules of inc logic}
\end{figure}

\begin{theoremEnd}{theorem}[Soundness of Incorrectness Logic Rules in \allFailTopKATs]\label{the: incorrectness logic fail rule sound}
  The rules in \Cref{fig: all proof rules of inc logic} are valid for any
  FailTopKAT \((\algebra{F}, \algebra{K}, \algebra{B})\),
  \(a, b, c \in \algebra{B}\) and \(p, q \in \algebra{F}\).
\end{theoremEnd}

\begin{proofEnd}
    All of the \(\okState\) cases is proved in~\Cref{the: incor logic rule sound ok},
    therefore we just prove all the \(\errState\) case.

    We list all the failure cases of these rules
    and remove all the rules without failure cases here,
    for convenience of the reader:
    \begin{mathpar}
        \inferrule[Empty]
        {\\}{\incorTriple{b}{p}{\errState: 0}}

        \and
        \inferrule[Consequence]
        {b \leq b' \\ \incorTriple{b}{p}{\errState: c} \\ c' \leq c}
        {\incorTriple{b'}{p}{\errState: c'}}

        \and
        \inferrule[Disjunction]
        {\incorTriple{b_1}{p}{\errState: c_1} \\ \incorTriple{b_2}{p}{\errState: c_2}}
        {\incorTriple{b_1 + b_2}{p}{\errState: (c_1 + c_2)}}

        \and
        \inferrule[Identity]
        {\\} {\incorTriple{b}{1}{\errState: 0}}

        \and
        \inferrule[Composition-Fail]
        {\incorTriple{a}{p}{\errState: b}}
        {\incorTriple{a}{p q}{\errState: b}}

        \and
        \inferrule[Composition-Normal]
        {\incorTriple{a}{p}{\okState: b} \\ \incorTriple{b}{p}{\errState: c}}
        {\incorTriple{a}{p}{\errState: c}}

        \and
        \inferrule[Choice-Left]
        {\incorTriple{b}{p}{\errState: c}}
        {\incorTriple{b}{p + q}{\errState: c}}

        \and
        \inferrule[Choice-Right]
        {\incorTriple{b}{q}{\errState: c}}
        {\incorTriple{b}{p + q}{\errState: c}}

        \and
        \inferrule[Assume]
        {\\}
        {\incorTriple{a}{b}{\errState: 0}}

        \and
        \inferrule[Error]
        {\\}{\incorTriple{b}{\fail}{\errState: b}}

        \and
        \inferrule[Iter-NonZero]
        {\incorTriple{b}{\starOf{p}p}{\errState: c}}
        {\incorTriple{b}{\starOf{p}}{\errState: c}}
    \end{mathpar}

    The proof are basically the same as~\Cref{the: incor logic rule sound ok},
    since \(\allFailTopKATs\) share most of the properties of \(\allTopKATs\):
    \begin{itemize}
        \item \itemTitle{Empty:}
            By left-annihilation and \(0\) is the least element (additive identity):
            \[\top b p \geq 0 = 0 \fail\]
        \item \itemTitle{Consequence Rule:}
            because multiplication preserves order
            and \(b' \geq b\), therefore \(\top b' p \geq \top b p\);
            and because \(c' \leq c\), then \(c' \fail \leq c \fail\).
            Therefore
            \[\top b' p \geq \top b p \geq c \fail \geq c' \fail\]
        \item \itemTitle{Disjunction Rule:}
            because addition preserves order, therefore
            \[\top b_1 p + \top b_2 p \geq c_1 \fail + c_2 \fail\]
            and by distributivity, we have
            \[\top (b_1 + b_2) p \geq (c_1 + c_2) \fail\]
        \item \itemTitle{Identity Rule:}
            By left-annihilation and \(0\) is the least element (additive identity):
            \[\top b 1 \geq 0 = 0 \fail\]
        \item \itemTitle{Composition-Fail Rule:}
            Given \(\top a p \geq b \fail\),
            then by order preserving and associativity of multiplication,
            we have
            \[\top a p q \geq b \fail q = b \fail\]
        \item \itemTitle{Composition-Normal Rule:}
            first, by idempotency of \(\top\)
            and associativity of multiplication, we have
            \[\top a (p q) = \top \top a p q\]
            By order preserving of multiplication,
            and the two premises \(\top a p \geq b\), \(\top b p \geq c \fail\),
            we have \[\top a (p q) = \top \top a p q \geq \top b q \geq c \fail\]
        \item \itemTitle{Choice-Left and Choice-Right Rule:}
            by distributivity, we have
            \[\top b (p + q)
            = \top b p + \top b q
            \geq \top b p \geq c \fail\]
            and
            \[\top b (p + q)
            = \top b p + \top b q
            \geq \top b q \geq c \fail\]
        \item \itemTitle{Error:}
            because \(\top \geq 1\), by order preserving of multiplication
            \[\top b \fail \geq b \fail\]
        \item \itemTitle{Iter-NonZero Rule:}
            since \(\starOf{p} = p \starOf{p} + 1\),
            we have
            \[\top b \starOf{p}
            = \top b (p \starOf{p} + 1)
            \geq \top b p \starOf{p} \geq c \fail\]
    \end{itemize}

    Note from the proof we can see all the rules with failure case \(\errState: 0\),
    like Assume and Identity rule,
    are redundant, since the failure case rule can be derived from the failure case of empty rule.
\end{proofEnd}

\section{Examples: Reasoning Using TopKAT and FailTopKAT}

In this section, we show some concrete examples of algebraic program reasoning.
We take the assignment language of \citet[Fig.~2]{ohearn_incorrectness_2020},
and regard assignments as primitive actions and assume statements as primitive tests.
The relational semantics of this language forms a relational FailTopKAT,
and it also forms a relational TopKAT if we do not consider the \(\comError\) command and
the \(\errState\) post-condition.

\begin{example}[Incorrect Absolute Value Procedure]\label{exp: incorrect absolute value}
    Here is an incorrect procedure for finding the absolute value of \(x\):
    \[\comITE{x < 0}{\comSkip}{\comAssign{x}{-x}},\]
    To have a correct procedure for computing the absolute value the condition of the if statement should be \(x > 0\).
    We can use TopKAT to show that every negative number is reachable by using 
    the following incorrectness triple: 
    \[\incorTriple{x < 0}{\comITE{x < 0}{\comSkip}{\comAssign{x}{-x}}}{x < 0}\]

This triple can be proven using just the theory of TopKAT. 
First, we can unfold the if statement:
\[\incorTriple{x < 0}{ (x < 0) 1 + \compl{(x < 0)} (\comAssign{x}{-x})}{x < 0}\]
Then convert the triple to TopKAT encoding: 
\[\top(x < 0) ((x < 0) 1 + \compl{(x < 0)} (\comAssign{x}{-x})) \geq (x < 0)\]
Finally, we prove the above inequality:
\begin{align*}
    & \top(x < 0) ((x < 0) 1 + \compl{(x < 0)} (\comAssign{x}{-x}))\\
    & \geq \top (x < 0) (x < 0) 1 
        & (p + q) \geq p\\
    & = \top (x < 0) 
        & \text{idempotency of test} \\
    & \geq (x < 0)  & \top \geq 1
\end{align*}
Thus we have shown that 
\[\incorTriple{x < 0}{ (x < 0) 1 + \compl{(x < 0)} (\comAssign{x}{-x})}{x < 0}\]
is valid, and the non-desirable results in \(x < 0\) can be reached.
As this triple can be shown just using the equational theory of TopKAT, 
this triple can be automatically decided using the algorithm in \Cref{the: PSPACE for TopKAT equalities}.
\end{example}     

\citeauthor{ohearn_incorrectness_2020} motivated the under-approximate triple 
as a way to reason about incorrect programs. 
However the under-approximation logic can have other use cases.
For example, we can mix under-approximation and over-approximation triples to
prove a certain post condition is the strongest
(as in Hoare logic) without a relational semantics. We show this next.

\begin{example}[Reasoning With Hoare And Incorrectness Logic]\label{exp: hoare and incorrect together}
    The assertion \(x \geq 0\) is the strongest post condition of program 
    \(\comWhile{x < 0}{\comAssign{x}{x + 1}}\) with precondition \(\true\).

    We can show this by the following two triples:
\begin{align*}
    & \incorTriple{\true}{\comWhile{x < 0}{\comAssign{x}{x + 1}}}{x \geq 0}; \\
    & \hoareTriple{\true}{\comWhile{x < 0}{\comAssign{x}{x + 1}}}{x \geq 0}.
\end{align*}

Because for all \(p\) in some TopKAT, \(\starOf{p} \geq 1\),
the incorrectness triple can be shown as follows
\[\top 1 \starOf{((x < 0) (\comAssign{x}{x + 1}))}\compl{(x < 0)}
\geq \top 1 1 \compl{(x < 0)}
\geq (x \geq 0).\]
And because \(\top \geq 1 \starOf{((x < 0) (\comAssign{x}{x + 1}))}\),
then the Hoare triple can be shown
\[
    1 \starOf{((x < 0) (\comAssign{x}{x + 1}))}\compl{(x < 0)}
    \leq \top \compl{(x < 0)}
    = \top (x \geq 0)
  \]
  \end{example}

\begin{example}[Theorem Proving In Hoare And Incorrectness Logic]
    For all TopKATs \((\algebra{K}, \algebra{B})\), and \(b, c \in \algebra{B}\), \(p \in \algebra{K}\),
    if \(c \geq \compl{b}\), then following incorrectness and Hoare triples are valid
    \[\incorTriple{c}{\comWhile{b}{p}}{\compl{b}} 
    \tand 
    \hoareTriple{c}{\comWhile{b}{p}}{\compl{b}}\]

  This example is a generalization of \Cref{exp: hoare and incorrect together}.
If we have a while loop with condition \(b\),
and the precondition \(c\) is larger than \(\compl{b}\),
then \(\compl{b}\) is the strongest post-condition (in the sense of Hoare logic).

We first show a proof in a relational setting, as a comparison to the algebraic proof.
\begin{itemize}
    \item \(\codomain{c \starOf{(b p)} \compl{b}} \supseteq \codomain{\compl{b}}\):
        because on the left hand side, every output needs to go through the final check 
        of \(\compl{b}\), it will also be in \(\codomain{\compl{b}}\);
    \item \(\codomain{c \starOf{(b p)} \compl{b}} \subseteq \codomain{\compl{b}}\):
        if the input of the left hand side is in \(\codomain{\compl{b}}\), 
        it will not be filtered out by \(c\), will not go into loop \(\starOf{(b p)}\),
        will not be filtered out by \(\compl{b}\), and will be outputted unchanged. 
        thus everything in \(\codomain{\compl{b}}\) will 
        be a output of \( \starOf{(b p)} \compl{b}\),
        hence in \(\codomain{c \starOf{(b p)} \compl{b}}\).
\end{itemize}

Alternatively, we prove this example algebraically.
Because \(\starOf{(bp)} \geq 1\) and \(c \geq \compl{b}\): 
\[\top c \starOf{(b p)} \compl{b} \geq \top c 1 \compl{b} = \top \compl{b} \geq \compl{b},\]
and because \(\top \geq \top c \starOf{(b p)}\):
\[\compl{b} \leq \top \compl{b} \geq \top c \starOf{(b p)} \compl{b}.\]
We can also show that the Hoare triple is also valid with Kozen's encoding:
\[c \starOf{(bp)} \compl{b} (\compl{\compl{b}}) = c \starOf{(bp)} \compl{b} b = c \starOf{(bp)} 0 = 0.\]
The fact that we can also use Kozen's encoding to reach the same conclusion is not surprising,
as we have shown that Kozen's encoding is equivalent to our encoding in all TopKAT.
 \end{example}

\begin{example}[Error In Loop]
    This example simulates a while loop where the body will encounter an error when \(x \leq 0\),
    and it will do some useful computation \(p\) if it does not encounter the error:
    \[
        \incorTriple{\true}
        {\comWhile{x \geq 0}{\comITE{x \leq 0}{\comError}{p}}}
        {\errState: x = 0}
    \]
    the incorrect loop condition will trigger the possible error in the loop body,
    which is undesirable.

To show this triple, we need to show the following FailTopKAT inequalities:
\[
    \top 1 \starOf{((x \geq 0) ((x \leq 0) (\fail) + (\compl{x \leq 0}) (p)))} (\compl{x \geq 0}) \geq 
    (x = 0) \fail
\]
The proof is as follow: 
\begin{align*}
    & \top 1 \starOf{((x \geq 0) ((x \leq 0) (\fail) + (\compl{x \leq 0}) (p)))} (\compl{x \geq 0})\\
    & \geq \top 1 \starOf{((x \geq 0) (x \leq 0) (\fail))} (\compl{x \geq 0})
        & q + r \geq q\\
    & = \top 1 (x \geq 0) (x \leq 0) (\fail) (\compl{x \geq 0})
        & \starOf{q} \geq q \\
    & = \top 1 (x \geq 0) (x \leq 0) (\fail) 
        & \fail p = \fail \\
    & = \top (x = 0) \fail 
        & (x \geq 0 \land x \leq 0) = (x = 0)\\
    & \geq (x = 0) \fail & \top \geq 1
\end{align*}

Notice that in the second last step we used the fact 
\[(x \geq 0 \land x \leq 0) = (x = 0).\]
We invoked the logical meaning of \(x \geq 0\), \(x \leq 0\), and \(x = 0\).
Thus we are \emph{not} purely using the theory of TopKAT.
\end{example}

We can use more than just logical implications.
Since our encoding of incorrectness logic is conservative (relationally valid), 
all the proof rules of incorrectness logic will play nicely with algebraic reasoning.
The next example demonstrates the mix of equational reasoning and assignment rule.

\begin{example}[Assignment]
    We have the same program as \Cref{exp: incorrect absolute value}
    with the precondition changed to \(x \geq 0\):
    \[\incorTriple{x > 0}{\comITE{x < 0}{\comSkip}{\comAssign{x}{-x}}}{x < 0}.\]

    Even though the assignment rule from Incorrectness Logic is not propositional, 
    in the sense we use in this paper,  
    we can use specific inequalities to represent specific instances of these rules. 
    For example, we can use an instance represented by the following incorrectness triple:
\[\incorTriple{x > 0}{\comAssign{x}{-x}}{x < 0},\]
This triple corresponds to the following TopKAT inequality:
\[\top (x > 0) (\comAssign{x}{-x}) \geq (x < 0).\]
With the above inequality, we can derive the incorrectness triple in the example as follow:
\begin{align*}
    & \top (x > 0) ((x < 0) 1 + \compl{(x < 0)} (\comAssign{x}{-x})) \\
    & \geq \top (x > 0) (\compl{(x < 0)} (\comAssign{x}{-x})) 
        & p + q \geq q\\
    & \geq \top (x > 0) (\comAssign{x}{-x}) 
        & (x > 0) \leq \compl{(x < 0)}\\
    & \geq (x < 0) 
        & \text{above inequality}
\end{align*}
\end{example}

Thus, as we can see, even though TopKAT does not contain an assignment axiom,
we can still reason about programs with assignments.

\section{Related Encodings in Kleene Algebras}\label{sec:related-encodings}
The use of the complete relation as the top element in a relation-based algebraic structure 
traces back to the study of relation algebra~\cite{andreka_axiomatizability_2011, maddux_origin_1991}.
However extending relational Kleene algebra with complete relation 
was only recently studied by \citet{pous_automata_2016,pous_kleene_2013}.
The counterexample for completeness provided by \citeauthor{pous_automata_2016} 
can also use to disprove completeness of relational TopKAT,
\[\top \synt{p} \top \synt{q} \top = \top \synt{q} \top \synt{p} \top.\]
However, our counterexamples \(\top \synt{p} = \top \synt{p} \top \synt{p}\)
and \(\synt{p} \top \synt{p} \geq \synt{p}\) are simpler than~\citeauthor{pous_automata_2016}'s.
We also notice that \(\synt{p} \top \synt{p} \geq \synt{p}\) is surprisingly similar 
to the counterexample for completeness of relational KAC\(^{-}\)
(Kleene algebra with converse)~\cite{hutchison_kleene_2014}, which is
\[\synt{p} \reverse{\synt{p}} \synt{p} \geq \synt{p}.\]
\citet{hutchison_kleene_2014} solved the incompleteness problem 
by extending the equational system of KAC\(^{-}\) with the counterexample 
\(\synt{p} \reverse{\synt{p}} \synt{p} \geq \synt{p}\) 
obtaining the relationally complete system KAC\@.

To further expand on the potential connections between TopKAT and KAC,
we have discovered that the converse relation may also be able to express codomain.
Our original formulation exploits the fact that for every relation \(p\), 
\(\reverse{p} p\) is larger than the identity relation on codomain of \(p\)
(\(\{(x, x) \mid x \in \codomain{p}\}\)),
and smaller than the complete relation on codomain of \(p\)
(\(\{(x, y) \mid x, y \in \codomain{p}\}\)).
Therefore, a relationally valid encoding for incorrectness triple \(\incorTriple{b}{p}{c}\) can be 
\[\reverse{p} b p \geq c.\]
several downsides of the KAT with converse encoding is the complicated equational theory of KAC
and the lack of an obvious way to formulate Hoare logic.
However, this discovery still shines a light on the connection of KAC and codomain.

Contrary to the aforementioned attempt to capture the naive codomain of relations,
there are other works that seek to have domain and codomain as a built-in operator in the algebraic theory
\cite{fahrenberg_domain_2021,desharnais_kleene_2006,desharnais_modal_2004}.

In a concurrent recent work, \citet{fahrenberg_algebra_2021} showed an encoding of incorrectness logic in an extension of 
Kleene Algebra with a modal operator: $\bra{p}b$ models the strongest postcondition of the program $p$ given the precondition $b$ as a test. 
As in our work, they can also encode both Hoare and incorrectness triples:
\begin{align*}
    \incorTriple{b}{p}{c} & \definedAs \bra{p} b \geq c,\\
    \hoareTriple{b}{p}{c} & \definedAs \bra{p} b \leq c.
\end{align*}
Their algebra is called CTC (Countably Test Complete) Modal Kleene Algebra, since they require all countable join of tests to exists, in order to obtain relative completeness of the incorrectness logic encoding.

\section{Related works}

\paragraph{Kleene Algebra with Tests and extensions}
The idea of Kleene Algebra with Tests was introduced by
\citet{kozen_kleene_1997} and its theory was studied in several
subsequent works.
\citet{goos_kleene_1997} focused on completeness and decidability
of KAT equalities. In particular, 
they showed that KAT is complete over relational models and language models, 
also deciding equality of KAT terms is PSPACE-complete by reduction to PDL\@.
\citet{cohen_complexity_1999} gave a more elementary proof of the PSPACE complexity of 
deciding equality in KAT\@.
In his seminal work, \citet{kozen_hoare_2000} showed that KAT subsumes
partial correctness of propositional Hoare logic.  This result
demonstrates the power of KAT in expressing program logics.

Inspired by earlier studies on relation algebra,
\citet{hutchison_kleene_2014} developed an extension of Kleene Algebra
with a converse relation. This system can also be used to express
incorrectness logic, however it has a more complex
equational theory than the one of TopKAT\@, which we present here.
An extension of the equational theory with a top element was also
considered in~\cite{pous_automata_2016} in the context of KA\@. This
work showed the incompleteness of this extension over relational
models. A top element was also used by~\citet{esparza_equational_2017}
as a way to ``forget the program state''. The same work also extended
KAT with failure to reason about abnormal termination.  In a different
direction, \citet{anderson_netkat_2014} extended KAT to NetKAT in
order to provide a semantical foundation of network applications; and
finally \cite{thiemann_probabilistic_2016} further extended NetKAT to
incorporate probabilistic reasoning, 
and \citet{smolka_cantor_2017} gives a new semantical foundation for ProbNetKAT.
\Citet{doumane_kleene_2019} gives a general way to extend Kleene Algebra
with a set of hypothesis, 
later \citet{pous_tools_2021} provide a way to derive 
completeness result for general extensions of Kleene Algebra.
In this paper, we use a more elementary proof similar to \citet{goos_kleene_1997} 
and \citet{cohen_complexity_1999},
instead of a more general approach as suggested by \citet{pous_tools_2021}.
KAT has also been integrated into a Coq library by
\citet{pous_kleene_2013}. This library can be used to
prove equivalences and correctness of while programs.

\paragraph{Incorrectness Logic and extensions}
Incorrectness logic has been recently introduced by
\citet{ohearn_incorrectness_2020} to reason about incorrect
programs. O'Hearn was motivated by the practical need of providing
proofs of failure and incorrectness. In his paper O'Hearn proposed a
proof system for incorrectness logic and studied its underlying semantics.
A similar system to the one studied by O'Hearn was investigated by
\citet{barthe_reverse_2011} for reasoning about randomized algorithms.
\citet{Murray_2020} implemented and 
formally verified a relational version of incorrectness logic in
Isabelle. \citeauthor{Murray_2020}'s logic is relational in the 
sense that it allows one to reason about two executions of two 
potentially different programs. It would be interesting to see if 
a similar logic could also be embedded naturally in TopKAT.
\citet{RaadBDDOV20} combined incorrectness logic and separation logic
to reason about incorrect programs in a local way without tracking the
global state.

As we discussed in the previous section, in a recent concurrent work \citet{fahrenberg_algebra_2021} showed an encoding of incorrectness logic in an extension of 
Kleene Algebra with a modal operator for representing strongest postconditions.
Our work differs from theirs in several aspects. 
First, we show the impossibility of encoding incorrectness logic in basic KAT. 
Second, we consider TopKAT rather than adding modal operators. 
The two approaches share some similarities, especially in relational models, 
where they are essentially equivalent, as discussed in \citet{fahrenberg_algebra_2021}. 
Third, we do not require all countable join of test to exist, 
since we don't focus on relative completeness. 
On the other hand, we study the meta-theory of TopKAT in detail.

\section{Conclusion and Future Work}\label{sec: future problem}


We believe that our work has clarified the main questions about how to
perform incorrectness reasoning in an equational algebraic system 
in the style of KAT\@. However, this work has also generated several
other interesting questions. We discuss some of them here.

\paragraph{Completeness.}
We have shown that the equational theory of TopKAT is incomplete over
relational TopKATs. This means that there might be valid incorrectness
triples in relational TopKAT that cannot be validated by the
equational theory of TopKAT\@.  Hence, a natural open question is to
investigate whether there are additional axioms that we could add to
recover completeness over relational TopKATs. One way to approach this
question is to consider an extension similar to the one studied by
\citet{hutchison_kleene_2014} that we discussed in
Section~\ref{sec:related-encodings}. It is natural to wonder if
extending TopKAT with a similar rule can help to recover completeness
over relational TopKATs.

\paragraph{Other directions.}
There are various abstractions of domain,
namely~\cite{fahrenberg_domain_2021,desharnais_kleene_2006,desharnais_modal_2004},
The conventional wisdom would suggest that these direct abstractions
are more powerful than TopKAT (admits more models), but it would also
be interesting to better understand the connections between the two.
%
\citet{fischer_propositional_1979, goos_kleene_1997}
showed strong connections between Kleene algebra with tests and 
propositional dynamic logic (PDL).
It would be interesting to see how TopKAT would relate to  
propositional dynamic logic.







\ifacm
\else 
\section*{Acknowledgements}

This material is based upon work supported by the
National Science Foundation under Award
No.~CNS 2040249 and Grant No.~CNS 2040222.  
Any opinions, findings, and
conclusions or recommendations expressed in this material are those
of the author and do not necessarily reflect the views of the
National Science Foundation.

We thank Damien Pous, Alexandra Silva, Bernhard M\"oller, Peter O'Hearn,
and all of our reviewers for their valuable inputs on this paper. 
And we thank Damien Pous and Jana Wagemaker for pointing 
out a mistake in a previous version of this paper.
\fi

\bibliography{IncorrectnessLogic.bib,KeleenAlgebraWithTest.bib,RelationAlgebra.bib,Other.bib}

\iffull

\appendix
\section{Appendix}

\begin{theoremEnd}[normal]{lemma}[redundancy of alphabet]\label{the: redundancy of alphabet}
    If for some \(K, B\) where \(\synt{p} \in K\) and \(\synt{b}, \synt{c} \in B\), 
    and a pair of terms \(\synt{t_1}, \synt{t_2}\),
    that for all relational KAT valuations \(u: K \union B \to \algebra{R}\),
    \[\validUnderVal{u}{\synt{t_1} = \synt{t_2}} \iff \validUnderVal{u}{\incorTriple{\synt{b}}{\synt{p}}{\synt{c}}}\]
    then there exist
    \(\hat{\synt{t_1}}, \hat{\synt{t_2}} \in \KATTerm{\{\synt{p}\}}{\{\synt{b},
        \synt{c}\}}\) such that, for every relational valuation
    \(\hat{u}: \{\synt{p}, \synt{b}, \synt{c}\} \to \algebra{R}\),
    \[\validUnderVal{\hat{u}}{\hat{\synt{t_1}} = \hat{\synt{t_2}}} \iff 
    \validUnderVal{\hat{u}}{\incorTriple{\synt{b}}{\synt{p}}{\synt{c}}}\]
\end{theoremEnd}

\begin{proofEnd}
    For every term \(\synt{t} \in \KATTerm{K}{B}\),
    we can construct \(\hat{\synt{t}} \in \KATTerm{\{p\}}{\{b, c\}}\)
    in the following way:
    \begin{itemize}
        \item Change all the primitive actions \(\synt{t}\) to \(\synt{p}\),
        \item for all primitive tests \(\synt{a}\) in \(\synt{t}\),
        if \(\synt{a} \not \in \{\synt{b}, \synt{c}\}\), 
        then we change \(\synt{a}\) into \(\synt{b}\) 
    \end{itemize} 

    for all relational KATs \(\algebra{R}\),
    given a valuation \(\hat{u}: \{\synt{p}, \synt{b}, \synt{c}\} \to \algebra{R}\),
    we can construct a valuation \(u: K \union B \to \algebra{R}\)
    as follows:
    \[u(\synt{p}) \definedAs
    \begin{cases}
    \hat{u}(\synt{p}) & \synt{p} \in K \\
    \hat{u}(\synt{c}) & \synt{p} = \synt{c} \\
    \hat{u}(\synt{b}) & \synt{p} \in B, \synt{p} \neq \synt{c} 
    \end{cases}\]

    Then by induction, we can have the following conclusion:
    for all relational KATs \(\algebra{R}\), 
    valuation \(\hat{u}: \{\synt{p}, \synt{b}, \synt{c}\} \to \algebra{R}\),
    and all \(\synt{t} \in \KATTerm{K}{B}\)
    \[\interpWith{\hat{u}}{\hat{\synt{t}}} = \interpWith{u}{\synt{t}}\]
    where the \(u\) and \(\hat{\synt{t}}\) are constructed 
    from \(\hat{u}\) and \(\synt{t}\) as described above.

    Thus, for every valuation 
    \(\hat{u}: \{\synt{p}, \synt{b}, \synt{c}\} \to \algebra{R}\),
    we can construct \(u : K \union B \to \algebra{R}\) as stated above,
    and because 
    \[\interpWith{\hat{u}}{\hat{\synt{t_1}}} = 
    \interpWith{u}{\synt{t_1}} \tand \interpWith{\hat{u}}{\hat{\synt{t_2}}} = 
    \interpWith{u}{\synt{t_2}}\]
    we have 
    \[\interpWith{\hat{u}}{\hat{\synt{t_1}}} = \interpWith{\hat{u}}{\hat{\synt{t_2}}} 
    \iff \interpWith{u}{\synt{t_1}} = \interpWith{u}{\synt{t_2}}\]
    By the premise, we can find a valuation \(u\) s.t.
    \(\validUnderVal{u}{\synt{t_1} = \synt{t_2}} \iff \validUnderVal{u}{\incorTriple{\synt{b}}{\synt{p}}{\synt{c}}}\)
    holds, therefore:
    \[\validUnderVal{\hat{u}}{\hat{\synt{t_1}} = \hat{\synt{t_2}}}
    \iff \interpWith{\hat{u}}{\hat{\synt{t_1}}} = \interpWith{\hat{u}}{\hat{\synt{t_2}}} 
    \iff \interpWith{u}{\synt{t_1}} = \interpWith{u}{\synt{t_2}}
    \iff \validUnderVal{u}{\incorTriple{\synt{b}}{\synt{p}}{\synt{c}}}\]
    Since \(u\) and \(\hat{u}\) agree on valuation value of \(\synt{b}, \synt{p}, \synt{c}\)
    therefore so is the generated interpretation \(\interpWith{u}{-}\) and \(\interpWith{\hat{u}}{-}\),
    hence we have:
    \[\validUnderVal{u}{\incorTriple{\synt{b}}{\synt{p}}{\synt{c}}}
    \iff \validUnderVal{\hat{u}}{\incorTriple{\synt{b}}{\synt{p}}{\synt{c}}}\]
    Finally, for all valuations \(\hat{u}\), we have:
    \[\validUnderVal{\hat{u}}{\hat{\synt{t_1}} = \hat{\synt{t_2}}}
    \iff \validUnderVal{\hat{u}}{\incorTriple{\synt{b}}{\synt{p}}{\synt{c}}}\]
\end{proofEnd}

Thus, in order to show that there does not exist a pair of terms
\(\synt{t_1}, \synt{t_2} \in \KATTerm{K}{B}\) to express incorrectness logic,
we only need to show that there does not exist
\(\hat{\synt{t_1}}, \hat{\synt{t_2}} \in \KATTerm{\{\synt{p}\}}{\{\synt{b},
    \synt{c}\}}\) that are capable of formulating incorrectness logic.

We first prove some property of interpretations:

\begin{theoremEnd}[normal]{lemma}[monotonicity of relational interpretation]\label{the: monotonicity of relational interpretation}
    For all terms \(\synt{t} \in \KATTerm{K}{B}\) and relational KAT valuations
    \(u, u': K \union B \to \algebra{R}\), if
\begin{align*}
    \forall \synt{p} \in K &, u'(\synt{p}) \supseteq u(\synt{p}) \\
    \forall \synt{b} \in B &, u'(\synt{b}) = u(\synt{b}),
\end{align*}
then
\[\interpWith{u'}{\synt{t}} \supseteq \interpWith{u}{\synt{t}}\]
\end{theoremEnd}

\begin{proofEnd}
By induction on the structure of \(\synt{t}\).
\end{proofEnd}

\begin{theoremEnd}[normal]{lemma}\label{the: only one action var}
    Given a term \(\synt{t} \in \KATTerm{\{\synt{p}\}}{\{\synt{b}, \synt{c}\}}\)
    and a relational KAT valuation \(u\) with \(u(\synt{p}) = \emptyset\), if
    \((x, y) \in \interpWith{u}{\synt{t}}\), we have \(x = y\).
\end{theoremEnd}

\begin{proofEnd}
By induction on the structure of \(\synt{t}\).
\end{proofEnd}

\begin{theoremEnd}[normal]{lemma}\label{the: KAT inccorrectness logic core lemma}
    Given a term \(\synt{t} \in \KATTerm{\{\synt{p}\}}{\{\synt{b}, \synt{c}\}}\)
    and a relational KAT valuation \(u\), if \((x, y) \in \interpWith{u}{\synt{t}}\),
    then either:
    \begin{itemize}
    \item \(x = y\) and \((x,y) \in \interpWith{u_{\emptyset}}{\synt{t}}\), with
    \(u_{\emptyset} \definedAs u[p \mapsto \emptyset]\); or
    \item there exist \(z\) and \(z'\) such that
    \((x, z) \in \interpWith{u}{\synt{p}} \tand (z', y) \in
    \interpWith{u}{\synt{p}}\).
\end{itemize}
\end{theoremEnd}

\begin{proofEnd}
    Intuitively, this lemma states
    if the element \((x, x)\) is generated by purely by some tests 
    in \(\interpWith{u}{\synt{t_2}}\),
    then we can ignore all the action variables in the term by setting it to \(\emptyset\).
    Otherwise, \((x, x)\) must be generated by composing some actions together,
    since tests only \emph{filters out} elements when composed 
    and cannot generated new elements.
    Thus we will need at least a action to start with \(x\), 
    and a action to end with \(x\).

    This lemma can be proven by induction on the structure of \(\synt{t}\):
    \begin{itemize}
        \item If \(\synt{t}\) is an element of the alphabet, i.e. \(\synt{t} = \synt{p}\), \(\synt{t} = \synt{b}\), or \(\synt{t} = \synt{c}\).
        \begin{itemize}
        \item If \(\synt{t} = \synt{p}\), then \((x, y) \in \interpWith{u}{\synt{t}}\) 
            iff \((x, y) \in \interpWith{u}{\synt{p}}\)
            thus, we can have \(z = y, z' = x\), hence 
            \[(x, z) \in \interpWith{u}{\synt{p}} \tand (z', y) \in \interpWith{u}{\synt{p}}\]
        \item If \(\synt{t} = \synt{b}\) or \(\synt{t} = \synt{c}\),
            then the valuation of \(\synt{p}\) will not matter, 
            hence assigning \(\synt{p}\) to empty will also contain \((x, y)\).
            Therefore \((x, y) \in \interpWith{u_{\emptyset}}{\synt{t}}\).
            And by \Cref{the: only one action var}, \(x = y\)
        \end{itemize}
        \item If \(\synt{t}\) is of the form \(\synt{t_1} + \synt{t_2}\),
        then 
        \begin{align*}
            (x, y) \in \interpWith{u}{\synt{t_1} + \synt{t_2}}
            & \implies (x, y) \in \interpWith{u}{\synt{t_1}} \union \interpWith{u}{\synt{t_2}} \\
            & \implies (x, y) \in \interpWith{u}{\synt{t_1}} \text{ or } (x, y) \in \interpWith{u}{\synt{t_2}}
        \end{align*}
        Without loss of generality, assume \((x, y) \in \interpWith{u}{\synt{t_1}}\).
        Then by induction hypothesis, we have 
        \begin{itemize}[nosep]
            \item either \(\interpWith{u_{\emptyset}}{\synt{t_1}}\) and \(x = y\)
            \item or \(\exists z, z', \tst (x, z), (z', y) \in \interpWith{u}{\synt{p}}\)
        \end{itemize}
        Because \(\interpWith{u_{\emptyset}}{\synt{t}} = 
        \interpWith{u_{\emptyset}}{\synt{t_1} + \synt{t_2}} = 
        \interpWith{u_{\emptyset}}{\synt{t_1}} \union \interpWith{u_{\emptyset}}{\synt{t_2}}\),
        therefore 
        \[(x, y) \in \interpWith{u_{\emptyset}}{\synt{t_1}} \implies (x, y) \in \interpWith{u_{\emptyset}}{\synt{t}}\]
        Thus we have
        \begin{itemize}[nosep]
            \item either \((x, y) \in \interpWith{u_{\emptyset}}{\synt{t}}\) and \(x = y\)
            \item or \(\exists z, z', \tst (x, z), (z', y) \in \interpWith{u}{\synt{p}}\)
        \end{itemize}
        \item If \(\synt{t}\) is of the form \(\synt{t_1} \synt{t_2}\),
        then 
        \[
            (x, y) \in \interpWith{u}{\synt{\synt{t_1}} \synt{t_2}}
            \implies \exists k, \tst (x, k) \in \interpWith{u}{\synt{t_1}} \tand (k, y) \in \interpWith{u}{\synt{t_2}} 
        \]
        Then by induction hypothesis for \(\synt{t_1}\), we have 
        \begin{itemize}[nosep]
            \item either \((x, k) \in \interpWith{u_{\emptyset}}{\synt{t_1}}\) and \(x = k\)
            \item or \(\exists z, z', \tst (x, z), (z', k) \in \interpWith{u}{\synt{p}}\)
        \end{itemize}
        by induction hypothesis for \(\synt{t_2}\)
        \begin{itemize}[nosep]
            \item either \((k, y) \in \interpWith{u_{\emptyset}}{\synt{t_2}}\) and \(k = y\)
            \item or \(\exists z, z', \tst (k, z), (z', y) \in \interpWith{u}{\synt{p}}\)
        \end{itemize}
        
        Then there are 4 different cases:
        \begin{itemize}
            \item If both 
            \begin{itemize}[nosep]
                \item \((x, k) \in \interpWith{u_{\emptyset}}{\synt{t_1}}\) and \(x = k\)
                \item \((k, y) \in \interpWith{u_{\emptyset}}{\synt{t_2}}\) and \(k = y\)
            \end{itemize}
            are true, then \((x, y) \in \interpWith{u_{\emptyset}}{t}\) by rule of composition;
            and \(x = y\) by transitivity of equality.
            \item if both 
            \begin{itemize}[nosep]
                \item \((x, k) \in \interpWith{u_{\emptyset}}{\synt{t_1}}\) and \(x = k\)
                \item \(\exists z, z', \tst (k, z), (z', y) \in \interpWith{u}{\synt{p}}\)
            \end{itemize}
            are true, because \(x = k\), thus from the second point we have 
            \[\exists z, z', \tst (x, z), (z', y) \in \interpWith{u}{\synt{p}}\]
            hence the result is true.
            \item if both 
            \begin{itemize}[nosep]
                \item \(\exists z, z', \tst (x, z), (z', k) \in \interpWith{u}{\synt{p}}\)
                \item \((k, y) \in \interpWith{u_{\emptyset}}{\synt{t_2}}\) and \(k = y\)
            \end{itemize}
            are true, because \(k = y\), thus from the first point we have 
            \[\exists z, z', \tst (x, z), (z', y) \in \interpWith{u}{\synt{p}}\]
            hence the result is true.
            \item if both 
            \begin{itemize}[nosep]
                \item \(\exists z, z', \tst (x, z), (z', k) \in \interpWith{u}{\synt{p}}\)
                \item \(\exists z, z', \tst (k, z), (z', y) \in \interpWith{u}{\synt{p}}\)
            \end{itemize}
            is true, then there exists \((x, z)\) and \((z', y)\) in \(\interpWith{u}{\synt{p}}\)
            hence the result is valid
        \end{itemize}
        \item If \(\synt{t}\) is of the form \(\starOf{\synt{t_1}}\) for some \(\synt{t_1}\).
        Then by definition of \(\starOf{-}\) operator in \(\allRELs\),
        \((x, y) \in \starOf{\synt{t_1}}\) means there exists \(n \in \nat\),
        s.t. \((x, y) \in {(\synt{t_1})}^{n}\).
        Then we can prove this result by induction on \(n\), 
        using a strategy similar to the multiplication case.
        \item If \(\synt{t}\) is of the form \(\compl{\synt{t_1}}\),
        then by definition \(\synt{t_1}\) cannot contain primitive action.
        Hence \(\synt{t_1}\) and \(\synt{t}\) will not contain \(\synt{p}\).
        Therefore the valuation of \(p\) do not matter to the interpretation.
        Thus we have \[(x, y) \in \interpWith{u_{\emptyset}}{\synt{t}}\]
        and by \Cref{the: only one action var}, \(x = y\)
    \end{itemize}
\end{proofEnd}

\begin{theoremEnd}[normal]{lemma}[idempotency of top in \(\allTopKATs\)]\label{the: idempotency of top}
    In all TopKATs \[\top \top = \top\]
\end{theoremEnd}

\begin{proofEnd}
    First we show \(\top \top \leq \top\),
    by the axiom that \(\top\) is greater or equal to 
    all elements of the \(\allTopKATs\).

    Then we show \(\top \top \geq \top\).
    This is because \(\top \geq 1\), therefore
    \[\top \top \geq \top 1 \geq \top\]

    By anti-symmetry of ordering, we have \(\top \top = \top\)
\end{proofEnd}

\begin{definition*}[explicit definition of \(\allFailTopKATs\)]\label{def: explicit fail top kat}
    A \(\allFailTopKATs\) is an algebraic structure \((\algebra{F}, \algebra{K}, \algebra{B})\),
    where \(\algebra{F} \supsetneq \algebra{K} \supseteq \algebra{B}\),
    and \((\algebra{K}, \algebra{B})\) is a TopKAT
    (\(\algebra{F} \supsetneq \algebra{K} \) because 
    \(\algebra{F}\) is \(\algebra{K}\) with one new element \(\fail \not \in \algebra{K}\))
    For all \(p, q \in \algebra{F}\),  the following holds:
    \begin{align*}
        p  + 0 = 0 + p  & = p  & \text{identity} \\
        p  + q  &= q  + p  & \text{commutativity}\\
        (p  + q ) + r  & = p  + (q  + r ) & \text{associativity} \\
        p  + p  & = p  & \text{idempotent} \\
        1 p  = p  1 & = p  & \text{identity} \\
        0 p & = 0 & \text{left annihilation}\\
        (p   q )  r  & = p   (q   r ) & \text{associativity}\\
        (p  + q ) r  & = p r  + q r  & \text{distribution} \\
        r  (p  + q ) & = r p  + r q  & \text{distribution} \\
        \fail ~  p & = \fail & \text{failure} \\
        1 + (\starOf{p })p  = 1 + p (\starOf{p }) & = \starOf{p } 
            & \text{unfolding} \\
        q  + p r  \leq r  & \implies (\starOf{p })q  \leq r  & 
            \text{induction} \\
        q  + r p  \leq r  & \implies q (\starOf{p }) \leq r  & 
            \text{induction} \\
    \end{align*}
    where \(0, 1, \top\) are the additive identity,
    multiplicative identity, and top element in \(\algebra{K}\).
\end{definition*}

\begin{theoremEnd}[normal]{lemma}[multiplication and addition preserves order]\label{the: multiplication and addition preserves order in FailTopKAT}
    Given a \(\allFailTopKATs\) \(\algebra{F}\),
    For all \(p, q, r \in \algebra{F}\), if \(p \geq q\), then 
    \[
        p + r \geq q + r \tand
        p r \geq q r \tand
        r p \geq r q
    \]
\end{theoremEnd}

\begin{proofEnd}
Since \(p \geq q\), we have \(p + q = p\).

Therefore by associativity, commutativity, and idempotency of addition,
we have,
\[(p + r) + (q + r) = (p + q) + (r + r) = p + r\]
Hence \(p + r \geq q + r\).

By distributivity,
\[p r + q r = (p + q) r = pr\]
Hence \(pr \geq qr\).

By distributivity,
\[r p + r q = r (p + q) = r p\]
Hence \(rp \geq rq\).
\end{proofEnd}

\begin{theoremEnd}[normal]{lemma}[idempotency of top in \(\allFailTopKATs\)]
    In all FailTopKATs, \(\top \top = \top\)
\end{theoremEnd}

\begin{proof}
    same proof as \Cref{the: idempotency of top}
\end{proof}

\begin{theoremEnd}[normal]{lemma}[sum is \(\sup\)]\label{the:sum is sup in TopKAT}
    For all TopKAT \(\algebra{K}\) and finite set \(P \subseteq \algebra{K}\),    
    \[\sup P = \sum P,\]
    By distributivity, 
    \[\sup_{p \in P} (q \cdot p) = q \cdot (\sup_{p \in P} p)
    \tand \sup_{p \in P} (p \cdot q) = (\sup_{p \in P} p) \cdot q.\]
\end{theoremEnd}

\begin{proofEnd}
    \itemTitle{Base Case}: When \(P\) is empty, 
    then \(\sup P\) is the smallest element in the domain, hence \(0\);
    and \(\sum P\) is the additive identity, which is also \(0\).
    Thus \[\sup \emptyset = \sum \emptyset.\]

    \itemTitle{Induction Case}: 
    Assume \(\sup P = \sum P\) for all \(P\) of a certain length,
    then \(\sup (P \union \{p'\}) = \sum (P \union \{p'\})\), for all \(p' \notin P\).
    
    By induction hypothesis and \(p' \notin P\),
    \[\sum (P \union \{p'\}) = (\sum P) + p' = (\sup P) + p'.\]
    In order to show \((\sup P) + p' = \sup (P \union \{p'\}),\)
    
    \begin{itemize}
        \item we need to show for all \(p \in P \union \{p'\}\), \((\sup P) + p' \geq p\),
            this is easy:
            \[\begin{cases}
                (\sup P) + p' \geq (\sup P) \geq p & \text{if } p \in P \\
                (\sup P) + p' \geq p' = p & \text{if } p = p'
            \end{cases}\]
        \item Given another element \(q \geq p\) for all \(p \in P \union \{p'\}\),
            we need to show  \(q \geq (\sup P) + p'\).
            This statement can be shown by unfolding the definition of inequality and \(\sup\).
            
            Because \(q \geq p\) for all \(p \in P\), 
            thus \(q \geq \sup P\) and \(q + \sup P = q\);
            and because \(q \geq p'\), thus \(q + p' = q\).
            Thus \[q + (\sup P + p') = (q + \sup P) + p' = q + p' = q,\]
            we get \(q \geq (\sup P) + p'\).
    \end{itemize}
    Thus \(\sup (P \union \{p'\}) = \sum (P \union \{p'\})\).

    Finally, we have showed \((\sup P = \sum P)\) for all finite set \(P\).
\end{proofEnd}

\section{Proofs}

\printProofs{}

\fi
\end{document}